\documentclass[prl,twocolumn,superscriptaddress]{revtex4}
\usepackage{amsthm}
\usepackage{color}
\usepackage{amsmath,amsthm,mathrsfs,dsfont}
\usepackage{graphicx}
\usepackage{tikz}
\usepackage{xr}
\usepackage{xc}
\usepackage{float}
\usepackage{bbm}
\usepackage{epsfig} 
\usepackage{verbatim}
\usepackage{array}
\usepackage{setspace}
\usepackage{bbding}
\usepackage{amssymb}% http://ctan.org/pkg/amssymb
\usepackage{pifont}% http://ctan.org/pkg/pifont
\usepackage{braket}
\usepackage{tikz}
\usepackage{titlesec}

\usetikzlibrary{quantikz}

\usepackage{newtxtext,newtxmath}

\usepackage[unicode=true,
 bookmarks=true,bookmarksnumbered=false,bookmarksopen=false,
 breaklinks=false,pdfborder={0 0 1},backref=false,colorlinks=true]
 {hyperref}
\hypersetup{
 linkcolor=magenta, urlcolor=blue, citecolor=blue, pdfstartview={FitH}, hyperfootnotes=false, unicode=true}
 
\usepackage{color}
\newtheorem{lemma}{Lemma}
\newtheorem{theorem}{Theorem}

\newtheorem{definition}{Definition}
\newtheorem{problem}{Problem}

\makeatletter
\@ifundefined{textcolor}{}
{%
 \definecolor{BLACK}{gray}{0}
 \definecolor{WHITE}{gray}{1}
 \definecolor{RED}{rgb}{1,0,0}
 \definecolor{GREEN}{rgb}{0,1,0}
 \definecolor{BLUE}{rgb}{0,0,1}
 \definecolor{CYAN}{cmyk}{1,0,0,0}
 \definecolor{MAGENTA}{cmyk}{0,1,0,0}
 \definecolor{YELLOW}{cmyk}{0,0,1,0}
}

\usepackage{amsfonts}\usepackage{tabularx}\usepackage{dcolumn}\usepackage{bm}\usepackage{graphicx}\usepackage{epstopdf}

\hypersetup{urlcolor=blue}

\makeatother

\newcolumntype{C}[1]{>{\centering\arraybackslash$}p{#1}<{$}}

\usepackage{algorithm}
\usepackage{algorithmic}

\begin{document}

\widetext
\title{Exponential quantum advantages for practical non-Hermitian eigenproblems}

\author{Xiao-Ming Zhang}
%\affiliation{School of Physics, South China Normal University, Guangzhou 510006, China}
\affiliation {Key Laboratory of Atomic and Subatomic Structure and Quantum Control (Ministry of Education), Guangdong Basic Research Center of Excellence for Structure and Fundamental Interactions of Matter, School of Physics, South China Normal University, Guangzhou 510006, China} 

\affiliation {Guangdong Provincial Key Laboratory of Quantum Engineering and Quantum Materials, Guangdong-Hong Kong Joint Laboratory of Quantum Matter, South China Normal University, Guangzhou 510006, China}
\affiliation{Center on Frontiers of Computing Studies, School of Computer Science, Peking University, Beijing 100871, China}

\author{Yukun Zhang}
\affiliation{Center on Frontiers of Computing Studies, School of Computer Science, Peking University, Beijing 100871, China}

\author{Wenhao He}
\affiliation{Center for Computational Science and Engineering, Massachusetts Institute of Technology, Cambridge, MA 02139, USA}
\affiliation{School of Physics, Peking University, Beijing 100871, China}

\author{Xiao Yuan}
\email{xiaoyuan@pku.edu.cn}
\affiliation{Center on Frontiers of Computing Studies, School of Computer Science, Peking University, Beijing 100871, China}

\begin{abstract}
Non-Hermitian physics has emerged as a rich field of study, with applications ranging from $PT$-symmetry breaking and skin effects to non-Hermitian topological phase transitions. Yet most studies remain restricted to small-scale or classically tractable systems. While quantum computing has shown strong performance in Hermitian eigenproblems, its extension to the non-Hermitian regime remains largely unexplored. Here, we develop a quantum algorithm to address general non-Hermitian eigenvalue problems, specifically targeting eigenvalues near a given line in the complex plane---thereby generalizing previous results on ground state energy and spectral gap estimation for Hermitian matrices. Our method combines a fuzzy quantum eigenvalue detector with a divide-and-conquer strategy to efficiently isolate relevant eigenvalues. This yields a provable exponential quantum speedup for non-Hermitian eigenproblems. Furthermore, we discuss the broad applications in detecting spontaneous $PT$-symmetry breaking, estimating Liouvillian gaps, and analyzing classical Markov processes. These results highlight the potential of quantum algorithms in tackling challenging problems across quantum physics and beyond.

\end{abstract}
\maketitle

Interest in non-Hermitian many-body physics has grown rapidly in recent years, driven by the discovery of exotic phenomena such as spontaneous $PT$-symmetry breaking~\cite{Bender.98,Bender.99,Khare.00,Delabaere.00,Mostafazadeh.02,Mostafazadeh.02_,Ganainy.18,Wu.19,Weidemann.22}, non-Hermitian skin effects~\cite{Xiujuan.22}, and topology-driven phase transitions~\cite{Okuma.19,Longhi.19,Weidemann.22}, among others~\cite{Xiujuan.22,Okuma.19,Longhi.19,Cao.15,el.18,Ashida.20,Okuma.23}. 
From a theoretical {standpoint}, non-Hermitian physics naturally arises in open quantum systems~\cite{Dalibard.92,Carmichael.93,Nakagawa.18,Song.19}, weak measurement protocols~\cite{Dressel.14}, and Floquet engineering~\cite{Bukov.15,Oka.19}. 
{Experimentally}, non-Hermitian Hamiltonians have been implemented across various platforms, including solid-state systems~\cite{Wu.19}, photonic structures~\cite{Weidemann.22}, and other engineered setups~\cite{Ganainy.18}.
A central challenge in studying these systems lies in accurately characterizing their eigenvalue spectra. For instance, the relaxation dynamics of open quantum systems governed by master equations are determined by the Liouvillian gap i.e., the difference between the leading real parts of eigenvalues of the Lindblad superoperator. Similarly, the presence or absence of $PT$-symmetry in a non-Hermitian Hamiltonian is determined by whether the eigenvalues {are real or complex.}

Despite growing theoretical interest, current studies remain limited to small-scale or analytically tractable models. For general non-Hermitian many-body systems, the exponential growth of Hilbert space makes classical computation prohibitively expensive.
While quantum algorithms have demonstrated significant advantages in solving Hermitian problems~\cite{Poulin.09,Peruzzo.14,Somma.19,Ge.19,Lin.20,Zeng.21,Gharibian.22,Albash.18}, their extension to the non-Hermitian regime {faces} fundamental challenges. Quantum mechanics is inherently Hermitian, {hindering direct adaptation} of standard quantum techniques to non-Hermitian problems. Existing quantum approaches either rely on strong assumptions~\cite{Shao.19,Endo.20,Yoshioka.20,Xie.23,zhao.23,Low.24,Alase.24}—e.g., real eigenvalues for diagonalizable matrices~\cite{Shao.19}—or suffer from limited generality and uncertain performance, as in variational methods~\cite{Endo.20,Yoshioka.20,Xie.23,zhao.23}. In practice, non-Hermitian many-body systems often involve complex and even defective eigenvalues, especially in $PT$-broken phases. Thus, the development of efficient and broadly applicable quantum algorithms for non-Hermitian eigenproblems remains an open and critical {problem}.

\begin{figure} 
	\centering
	\includegraphics[width=.45\textwidth]{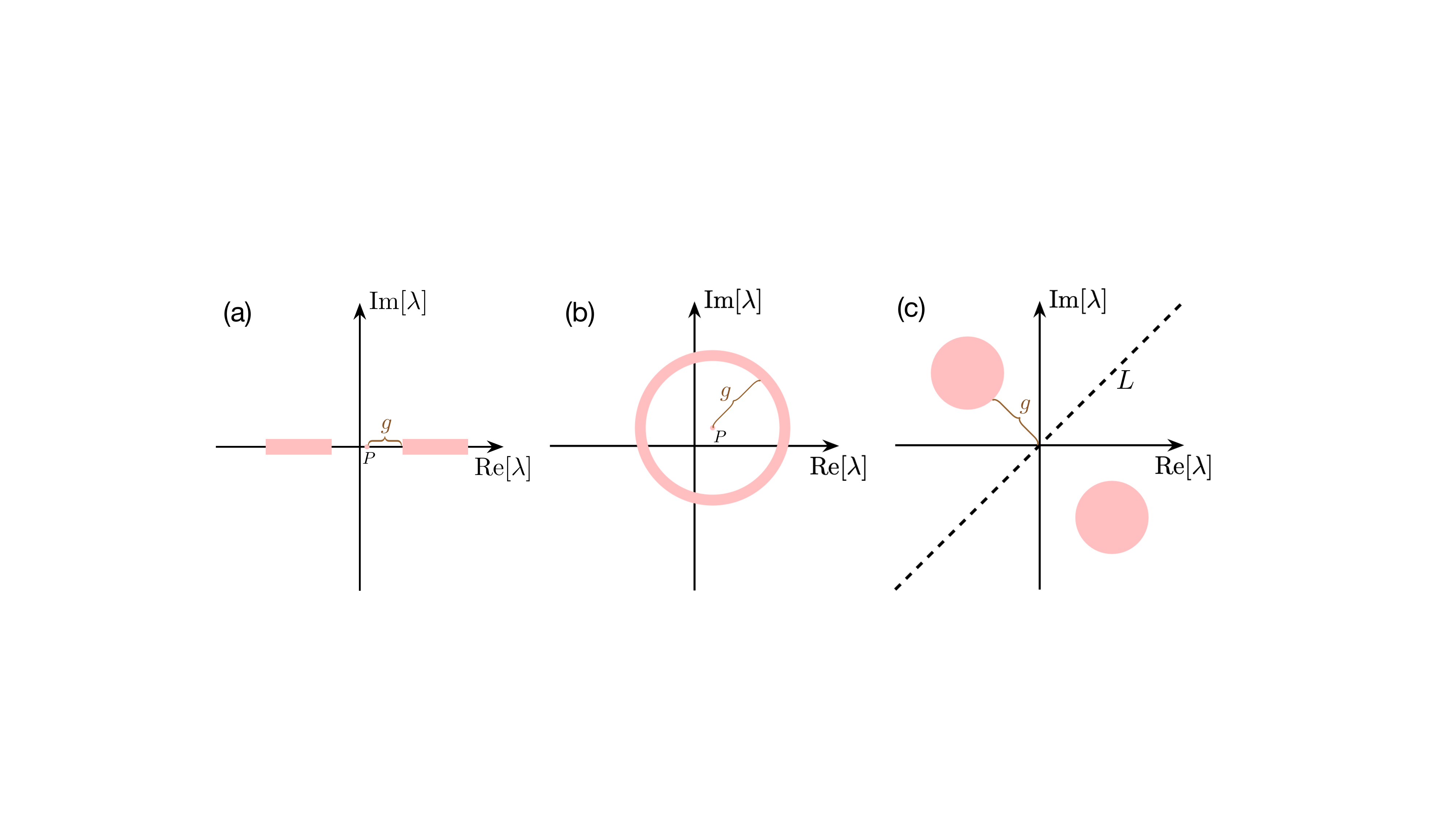} 
	\caption{Illustration of energy gaps.
		 (a) The energy gap for Hermitian matrices with real reference point $\boldsymbol{P}$. (b) Point gap for non-Hermitian matrices with complex reference point $\boldsymbol{P}$. (c) The line gap for non-Hermitian matrices with reference line $\boldsymbol{L}$.}
	\label{fig:gap} 
\end{figure}

Here, we propose an efficient quantum eigensolver for general matrices. As a representative example, we focus on the line gap problem---identifying eigenvalues closest to a given reference line in the complex plane (Fig.\ref{fig:gap}(c)). This framework can be naturally extended to other spectral tasks, such as the point gap problem (Fig.\ref{fig:gap}(b)). Our approach is built upon two key components: a fuzzy quantum eigenvalue detector {which identifies nearby eigenvalues}, and a divide-and-conquer strategy to efficiently locate them.
We further demonstrate that the decision version of the non-Hermitian eigenvalue problem is \textit{bounded-error quantum polynomial} (BQP)-complete, implying that, unless quantum computing can be efficiently simulated classically, no classical algorithm can solve this problem efficiently. This result provides strong theoretical evidence for the exponential quantum advantage of our algorithm. We also discuss several important applications, including estimating the Liouvillian gap in open quantum systems, detecting spontaneous symmetry breaking, and computing relaxation times in Markov processes---highlighting the broad utility of our method in non-Hermitian physics and beyond.

\vspace{0.2cm}
\noindent\textit{\textbf{Eigenvalue Problems.}} 
Given a square matrix $A \in \mathbb{C}^{N \times N}$,  complex eigenvalues $\lambda_j$ and the corresponding normalized eigenvectors $|v_j\rangle$ {are defined by}   
\begin{align}\label{eq:eig}
A|v_j\rangle = \lambda_j |v_j\rangle.
\end{align}
{Eigenvalue problems play fundamental} roles across virtually all fields of modern science and engineering.

In the Hermitian case, typical problems involve determining the ground-state energy (i.e., the smallest eigenvalue) or the energy gap between adjacent eigenvalues~\cite{Albash.18,Lin.20,Peruzzo.14}, as illustrated in Fig.~\ref{fig:gap}(a). However, for non-Hermitian systems, the concept of eigenvalue gaps is not unique~\cite{Kawabata.19,Borgnia.20,Bergholtz.21}. Depending on the physical context, one may define the gap with respect to a reference point or a reference line in the complex plane, as shown in Figs.~\ref{fig:gap}(b) and \ref{fig:gap}(c). These two types of gaps—point gaps and line gaps—are fundamentally distinct. 
For instance, in Fig.~\ref{fig:gap}(b), if a reference point $\bm{P}$ lies within the eigenvalue ring, and a reference line $\bm{L}$ passes through $\bm{P}$, the minimum distance from the eigenvalues to $\bm{P}$ remains nonzero, indicating a nonvanishing point gap. In contrast, the minimum distance to $\bm{L}$ becomes zero, indicating the absence of a line gap. Systems with nonzero point gaps and zero line gaps can naturally emerge due to differences in symmetry and topological characteristics~\cite{Kawabata.19,Borgnia.20,Bergholtz.21}.

From a computational standpoint, however, the methods for addressing point gap and line gap problems are largely similar. For this reason, we primarily focus on the line gap formulation in the main text.

\begin{problem}\label{prob:lg}
 Given a diagonalizable matrix $A\in\mathbb{C}^{N\times N}$ with spectral norm $\|A\|\leqslant1$, and a reference line $\boldsymbol{L}$. The goal is to output an eigenvalue $\lambda_{\min}$ closest to $\boldsymbol{L}$ up to an accuracy $\varepsilon\in(0,1)$, with promised that the distance from $\lambda_{\min}$ to $\boldsymbol{L}$ is larger than $\varepsilon$.
\end{problem}

\noindent Beyond its role in characterizing topology and symmetry, Problem~\ref{prob:lg} also corresponds to the evaluation of the \textit{Liouvillian gap} in the context of Lindbladian master equations~\cite{Medvedyeva.16,Banchi.17,Rowlands.18,Mori.20,Yuan.21,Zhou.22}, where the reference line $\bm{L}$ is taken as the imaginary axis. {The Liouvillian gap is central to dissipation dynamics, as it directly determines the system’s relaxation time.} Further discussions and applications of this concept will be presented {later}.

Note that when $\|A\| > 1$, one can always rescale the matrix by dividing it by a constant factor, such that the spectral norm of the resulting matrix becomes smaller than 1. The extension of our analysis to non-diagonalizable matrices is possible and is discussed in detail in~\cite{sm}. 
To simplify the discussion, we set $\bm{L}$ to be the real axis throughout this work, and assume that $\text{Im}(\lambda_j) > 0$. Solutions for other choices of $\bm{L}$ are structurally equivalent, and removing the $\text{Im}(\lambda_j) > 0$ restriction would only double the computational cost. Under this setting, the problem reduces to estimating the eigenvalue with the smallest imaginary part, which we denote as $\lambda_{\min} \equiv h + ig$. 
With regard to the allowable estimation error, the algorithm is considered successful if it returns a good approximation $\lambda_{\min}'$ of some eigenvalue such that $\text{Im}(\lambda_{\min}') - g \leq \varepsilon$.

We now proceed to introduce our main results.

\begin{figure} 
	\centering
	\includegraphics[width=.48\textwidth]{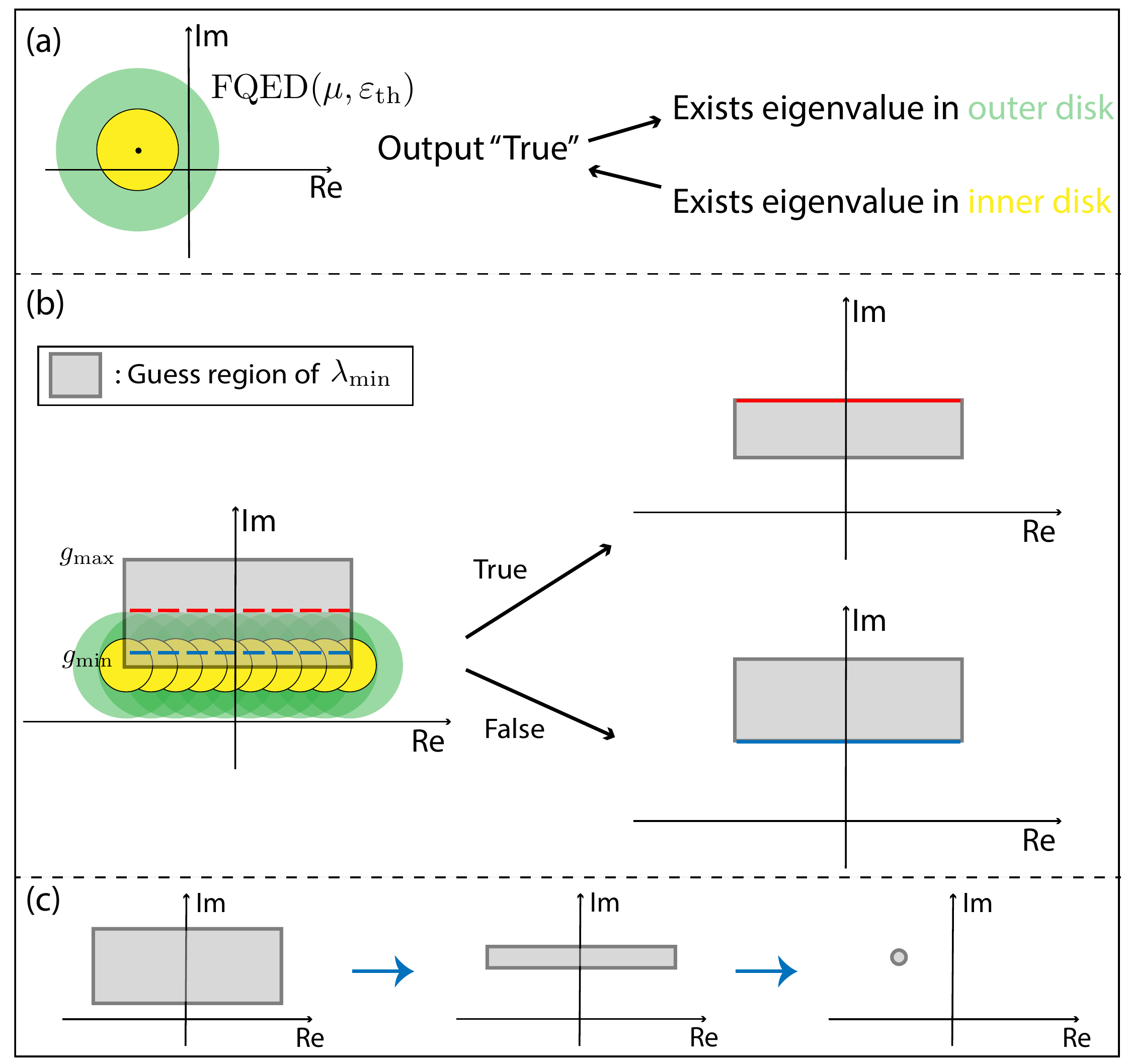} 
	\caption{(a) sketch of FQED$(\mu,\varepsilon_{\text{th}})$. Inner disk $\mathcal{D}(\mu,\varepsilon_{\text{th}})$ and outer disk $\mathcal{D}(\mu,2K\varepsilon_{\text{th}})$  are marked with yellow and green colors. (b) Each iteration for solving Problem~\ref{prob:lg}. Inner disks cover the line represented by $g_{\min}$, while the outer disks do not overlap with the real axis. If at least one of the FQEDs returns \texttt{True}, we update $g_{\max}$, otherwise we update $g_{\min}$.  (c) The guess region of the minimum eigenvalue shrinks iteratively until it is sufficiently small. In the last step, we output $\lambda_{\min}$ as the center of the last FQED returning \texttt{True}.}
	\label{fig:alg0} 
\end{figure}

\vspace{.2cm}
\noindent\textbf{\textit{Fuzzy quantum eigenvalue detector.}}
To begin with, we introduce the fuzzy quantum eigenvalue detector (FQED), as illustrated in Fig.~\ref{fig:alg0}(a). The FQED  takes two inputs: a complex center $\mu \in \mathbb{C}$ with $|\mu| < 1$, and a threshold parameter $\varepsilon_{\text{th}} \in (0,0.5)$. The output of the detector is a Boolean value, denoted as $\text{FQED}(\mu, \varepsilon_{\text{th}}) \in \{\texttt{True}, \texttt{False}\}$. 
With high probability (no greater than a small constant $\delta$), the output of FQED satisfies the following two properties:
\begin{itemize}
    \item[(1)] If $\text{FQED}(\mu, \varepsilon_{\text{th}})$ returns $ \texttt{True}$, then there exists at least one eigenvalue $\lambda_j$ such that 
    $
    \min_{\lambda_j} |\lambda_j - \mu| \leqslant 2K \varepsilon_{\text{th}}$,
    i.e., within the \textit{outer disk} $\mathcal{D}(\mu, 2K \varepsilon_{\text{th}})$.
    
    \item[(2)] If there exists an eigenvalue such that $
    \min_{\lambda_j} |\lambda_j - \mu| \leqslant \varepsilon_{\text{th}}$,
    i.e., within the \textit{inner disk} $\mathcal{D}(\mu, \varepsilon_{\text{th}})$, then $\text{FQED}(\mu, \varepsilon_{\text{th}})$  will return $\texttt{True}$.
\end{itemize}
Here, we define the disk $\mathcal{D}(a, b) = \{ x \in \mathbb{C} \,|\, |x - a| \leq b \}$, and the factor $K$ will be specified later. 
%The term “fuzzy” is borrowed from~\cite{Dong.22} for fuzzy bisection problem. 
When $ \min_{\lambda_j} |\lambda_j - \mu| \in(0,\varepsilon]$ or $ \min_{\lambda_j} |\lambda_j - \mu| \in(2K\varepsilon,\infty]$, FQED outputs $\texttt{True}$ or $\texttt{False}$, respectively. 
The term “fuzzy” highlights that when
$ \min_{\lambda_j} |\lambda_j - \mu| \in(\varepsilon,2K\varepsilon]$, the output could be arbitrary. This property arises from both the non-Hermiticity of the matrix $A$ and the finite-order truncation involved in the algorithm’s implementation.

To construct FQED, the first step is to reformulate the eigenvalue detection problem as a \textit{singular-value threshold} (SVT) problem. Consider the Jordan decomposition of the matrix $A$ given by
$A = P \Lambda P^{-1}$,
where $\Lambda$ is the Jordan canonical form (JCF) of $A$. We define the \textit{Jordan condition number} as $\kappa = \|P\| \cdot \|P^{-1}\|$\footnote{Due to the non-uniqueness of the Jordan decomposition, $\kappa$ is defined as the minimum over all valid decompositions.} which serves as a standard measure of eigenvalue sensitivity~\cite{Horn.12}. We require knowledge of an upper bound estimation of $\kappa$, denoted as $K$.  
We then define the function $C(\mu) \equiv \sigma_{\min}(\tilde{A})$, where $\tilde{A} \equiv A - \mu I$, and $\sigma_{\min}(\cdot)$ denotes the smallest singular value of the shifted matrix. A key observation is the following inequalities (see proof in the End Matter):
\begin{align}\label{eq:cu}
C(\mu) \leq \min_{\lambda_j} |\mu - \lambda_j| \leq K C(\mu).
\end{align}
We try to set $\text{FQED}(\mu, \varepsilon_{\text{th}}) = \texttt{True}$ when $C(\mu) \leq \varepsilon_{\text{th}}$, and $\text{FQED}(\mu, \varepsilon_{\text{th}}) = \texttt{False}$ when $C(\mu) > 2\varepsilon_{\text{th}}$. Then, because $\min_{\lambda_j}|\lambda_j-\mu|\leq\varepsilon_{\text{th}}$ implies $C(\mu)\leq\varepsilon_{\text{th}}$, property (2) is met obviously. Moreover, when $\min_{\lambda_j}|\lambda_j-\mu|>2K\varepsilon_{\text{th}}$, we have $C(\mu)> 2\varepsilon_{\text{th}}$, thus $\text{FQED}(\mu, \varepsilon_{\text{th}}) = \texttt{False}$. The contrapositive of this argument is equivalent to property (1). 
Consequently, the FQED reduces to an SVT problem, i.e. determining $C(\mu)\leqslant\varepsilon_{\text{th}}$ or $C(\mu)>2\varepsilon_{\text{th}}$. It can be efficiently addressed by exploiting the quantum singular value transformation (QSVT) framework~\cite{Gilyen.19}.

The application of QSVT requires a block encoding of the matrix $A$, i.e. a quantum circuit $\mathscr{O}_{A}$ applied at $(n+n_{\text{anc}})$ qubit (with $2^n=N$) satisfying
$
{A} = \left( \langle 0^{n_{\text{anc}}} | \otimes I \right) \mathscr{O}_{A} \left( |0^{n_{\text{anc}}} \rangle \otimes I \right)$, where $I$ is the $N$-dimensional identity operator. $\mathscr{O}_A$ can be efficiently constructed for a broad class of many-body quantum systems, up to a normalization factor~\cite{Childs.12}. %See also End Matter for illustration.
Moreover, the QSVT framework assumes the availability of a suitably prepared initial state—a standard assumption shared with conventional quantum phase estimation~\cite{Nielsen.02} and various projection-based Hermitian eigensolvers~\cite{Poulin.09,Peruzzo.14,Somma.19,Ge.19,Lin.20,Zeng.21,Gharibian.22}.
Let $\{\tilde{\sigma}_j, |v_j\rangle\}$ denote the singular value–right singular vector pairs of $\tilde{A}$, ordered by increasing singular value. %For initial state $|\psi_{\text{ini}}\rangle$, we define $\Pi^{(\tilde{A})}_{\varepsilon_{{\rm th}}}=\sum_{\tilde{\sigma}_{j}\leqslant \varepsilon_{\text{th}}}|v_j\rangle\langle v_j|$, and let $\tilde{\gamma}=\left\| \Pi^{(\tilde{A})}_{\varepsilon_{\text{th}}} |\psi_{\text{ini}}\rangle\right\|$ be the overlap to the low-singular-value subspace. 
Let $\Pi^{(\tilde{A})}_{\varepsilon_{{\rm th}}}=\sum_{\tilde{\sigma}_{j}\leqslant \varepsilon_{\text{th}}}|v_j\rangle\langle v_j|$ be the projector onto the low-singular-value subspace.
We require the knowledge of an overlap lower bound $\gamma$, such that the initial state $|\psi_{\text{ini}}\rangle$ satisfies:
\begin{align}\label{eq:overlap}
\left\| \Pi^{(\tilde{A})}_{\varepsilon_{\text{th}}} |\psi_{\text{ini}}\rangle \right\| \geq \gamma.
\end{align}
In practice, $|\psi_{\text{ini}}\rangle$ can be prepared using heuristic approaches such as variational quantum algorithms~\cite{Endo.20,Yoshioka.20,Xie.23,zhao.23} or adiabatic state preparation~\cite{Albash.18}. Alternatively, one may consider constructing an approximate model of $A$ whose eigenstructure is amenable to efficient classical computation.

With $\mathscr{O}_{A}$ and $|\psi_{\text{ini}}\rangle$, we give a detailed solution to the SVT problem in End Matter. The main idea is to transform the singular values of $\tilde{A}$ by a filter function, such that $f(\sigma_{j})\approx1$ for $\sigma_{j}>2\varepsilon_{\text{th}}$, and $f(\sigma_{j})\approx0$ for $\sigma_{j}\leqslant\varepsilon_{\text{th}}$. The unnormalized output state after projection into the target subspace is proportional to $\left(\sum_{j}f(\sigma_j)|u_j\rangle\langle v_j|\right)|\psi_{\text{ini}}\rangle$, whose amplitude is either $\Omega(\gamma)$ or close to zero for \texttt{True} or \texttt{False} cases respectively. Using binary amplitude estimation~\cite{Dong.22}, we can then construct FQED with gate complexity 
\begin{align}\label{eq:gc0}
C_{\text{FQED}}=\tilde{\mathcal{O}}\left(\gamma^{-1}\left(\varepsilon_{\text{th}}^{-1}\left(C_{\text{be}}+n\right)+C_{\text{sp}}\right)\right),
\end{align}
where $\tilde{\mathcal{O}}$ hides logarithmic dependencies on $K$, $\varepsilon_{\text{th}}$, $\gamma$, and $\delta$. Here, $C_{\text{be}}$ and $C_{\text{sp}}$ denote the gate counts for implementing  $\mathscr{O}_{A}$ and preparing $|\psi_{\text{ini}}\rangle$ respectively.

\vspace{.2cm}
\noindent\textbf{\textit{Divide-and-conquer query of FQEDs.}}  The FQED module constitutes the core component of our quantum algorithm. A straightforward approach {to} Problem~\ref{prob:lg} is to perform a sufficiently large number of queries to FQEDs with threshold parameter $\varepsilon / 2K$. Yet, instead of this naive strategy, we introduce a carefully designed divide-and-conquer search procedure that significantly reduces the overall runtime.

Specifically, we initialize the guess region for the line gap as $g \in [\varepsilon, 1+\varepsilon]$, and iteratively shrink this range through successive refinement. In the first iteration, we examine the presence of eigenvalues near the horizontal line defined by $\{c \,|\, \mathrm{Im}(c) = \varepsilon\}$. Let $\Delta = \varepsilon / 2K$ and $N_{\Delta} = \lceil 1/\Delta \rceil$. We query $\text{FQED}(r + i\varepsilon, \Delta)$ for all $r \in \{0, \pm \Delta, \pm 2\Delta, \ldots, \pm N_{\Delta} \Delta\}$ (see Fig.~\ref{fig:alg0}(b) for illustration).
If any of the FQED instances returns \texttt{True}, at least one eigenvalue lies within a corresponding outer disk. We thus update the {guess region} to $g \in [\varepsilon, 2\varepsilon]$. Conversely, if all detectors return \texttt{False}, this implies that no eigenvalues are located within any of the inner disks, allowing us to tighten the {guess region} to $g \in [\varepsilon(1+1/4K), 1+\varepsilon]$.
For the remaining iterations, the procedure follows a similar logic. Suppose the current {guess region} is $g \in [g_{\min}, g_{\max}]$. We set $\Delta = \min\left\{g_{\min}/2K, (g_{\max} - g_{\min})/4K\right\}$ and $N_{\Delta} = \lceil 1/\Delta \rceil$, and again query $\text{FQED}(r + i g_{\min}, \Delta)$ for all $r$ in the same set as above.
If any detector returns \texttt{True}, we refine the {guess region} to $g \in [g_{\min}, g_{\min} + 2K\Delta]$; otherwise, the {guess region} is updated to $g \in [g_{\min} (1+ \Delta/4K), g_{\max}]$. This process continues until the gap width satisfies $g_{\max} - g_{\min} \leq \varepsilon$, at which point we output the estimator of $\lambda_{\min}$ as the center of the final FQED instance that returned \texttt{True}.

We now estimate the total gate complexity. In each iteration, the algorithm performs $\mathcal{O}(\Delta^{-1})$ queries to FQEDs, each with a threshold parameter $\Delta$. Using Eq.~\eqref{eq:gc0}, the gate complexity of a single iteration is $
\tilde{\mathcal{O}}\left(\Delta^{-1} \gamma^{-1} \left((C_{\text{be}} + n)\Delta^{-1} + C_{\text{sp}}\right)\right)$.
In worst case, all FQED instances return \texttt{False} until the final iteration, requiring a total of $\tilde{\mathcal{O}}(K)$ iterations. Since $\Delta \geq \varepsilon/K$, the overall gate complexity is bounded by
\[
\tilde{\mathcal{O}}\left(K^2 \varepsilon^{-1}\gamma^{-1} \left( K\varepsilon^{-1}(C_{\text{be}} + n) + C_{\text{sp}} \right)\right).
\]
Thus, given efficient state preparation and block encoding, the runtime of our {algorithm} scales polynomially with $n$. In contrast, the runtime of typical classical algorithms scales exponentially with $n$, owing to the exponential growth of the matrix dimension $N = 2^n$. This quantum advantage will be further substantiated through the discussion of BQP-completeness presented below.

\vspace{0.2cm}
\noindent \textbf{\textit{Classical hardness and quantum advantage.}} To begin with, we consider a decision version of the non-Hermitian eigenvalue problem---determining whether the entire eigenspectrum of $A$ lies on the real axis (within a tolerable accuracy $\varepsilon$), or whether some eigenvalues possess nonzero imaginary components---whose physics motivation will be celebrated later. 
% This decision problem has a well-motivated physical context, which we will elaborate on later.
Formally,  let $g_{\max}\equiv\max(|{\rm Im}\lambda_j|)$ be the maximum imaginary part of eigenvalues and $\mathcal{V}_{\text{complex}}=\left\{|v_j\rangle:|\text{Im}(\lambda_j)|\geqslant\varepsilon\right\}$ be eigenvectors with large imaginary part, our decision problem is defined as follows.

\begin{problem}\label{prob:bqp2}

% Given a diagonalizable matrix $A$ with the following promise:
% (i) $(\alpha,n_{{\rm anc}})$-block encoding of $A$ can be efficiently constructed, and its Jordan condition number is upper bounded by $K$; (ii) Either $g_{\max}\geqslant \varepsilon$ (complex spectrum) or $g_{\max}\leqslant \varepsilon/2$ (real spectrum) is satisfied; and  (iii) One can efficiently prepare a state $|\psi'_{{\rm ini}}\rangle$, such that in complex spectrum case, we have $\left\|\langle v|\psi'_{{\rm ini}}\rangle\right\|\geqslant 2\gamma$ for some $|v\rangle\in\mathscr{V}_{\text{complex}}$. 
% Suppose $\alpha, n_{{\rm anc}}, K, \varepsilon, \gamma=O({\rm poly}(n))$. The goal is to output \texttt{True} for the complex spectrum case and output \texttt{False} in the real spectrum case. 
Consider a diagonalizable matrix $A$ with (i) efficient block encoding and an upper-bounded Jordan condition number $K$. (ii) Assume access to a guiding state $|\psi'_{{\rm ini}}\rangle$ satisfying $\left\|\langle v|\psi'_{{\rm ini}}\rangle\right\|\geqslant 2\gamma$ for some $|v\rangle\in\mathscr{V}_{\text{complex}}$. 
(iii) The problem setting assumes a promise that either \( g_{\max} \geq \varepsilon \) (complex spectrum), or \( g_{\max} \leq \varepsilon/2 \) (real spectrum). Promised that $\gamma^{-1},\varepsilon^{-1}, K=O(\text{poly}(n))$, the {goal} is to determine which of these two cases holds: \( g_{\max} \geq \varepsilon \), or \( g_{\max} \leq \varepsilon/2 \).
\end{problem}

 The approach is based on reducing it to the guided local Hamiltonian (GLH) problem~\cite{Gharibian.22}. %~\footnote{A difference from Theorem 1.2 in~\cite{Gharibian.22} is that the initial state has a nontrivial overlap with a specific ground state, instead of the ground state subspace. This condition can still be satisfied due to Eq.~(10) in~\cite{Gharibian.22}.}
Consider a Hermitian matrix $H$ with (i) efficient block encoding and (ii) access to a guiding state $|\psi_{{\rm ini},H}\rangle$ satisfying $|\langle {\rm gnd}_{H}|\psi_{{\rm ini},H}\rangle|\geqslant\gamma_H$ for the ground state $\ket{\rm gnd}_{H}$. (iii) Suppose either $\lambda_{\min,H}\in[0,a_H]$ (case 1) or $\lambda_{\min,H}\in[b_{H},1]$ (case 2) is satisfied, the goal of GHL is to distinguish these two cases. GLH problem is proven to be BQP-hard (i.e.~classically hard unless the universal quantum circuit can be efficiently simulated on classical computers) when $\gamma_{H}$ and $b_H-a_H$ are polynomial with respect to $n$. We show that any instance of the GLH problem can be efficiently reduced to an instance of Problem~\ref{prob:bqp2} by an adaptation of Corollary 6.27 in~\cite{low2017}.

\begin{lemma}[]\label{lm:low16}
For $b_H-a_H=\Omega(1/\text{poly}(n))$, there exists a real $d$-degree polynomial $f(x)=\sum_{j=0}^d\alpha_jx^j$, such that:
\begin{align}\label{eq:sign}
\min_{x\in[0,a_H]\cup[b_H,1]}|f(x)-{\rm sign}(x-(a_H+b_H)/2)| \leqslant2/3,
\end{align}
 for some $\sum_j|\alpha_j|=O({\rm poly}(n))$ and $d=O({\rm poly}(n))$.
\end{lemma}

\noindent Here, $f(x)$ can be regarded as an approximation of the sign function. For reduction, we construct a non-Hermitian matrix:
\begin{align}\label{eq:defa}
A = i C^{-1} \left( I - f(H) \right),
\end{align}
where $C = 1 + \sum_j |\alpha_j|$. We then establish the feasibility of the promises in Problem~\ref{prob:bqp2} for the matrix $A$.

Given an efficient block encoding of $H$, it is possible to efficiently construct the block encoding of $H^{p}$~\cite{Gilyen.19}, and consequently the block encoding en of $A$. Furthermore, it can be verified that the Jordan condition number of $A$ satisfies $K = 1$. 
In addition, the eigenvectors of $H$ and $A$ are identical, which implies that the initial state $|\psi_{\text{ini},H}\rangle$ naturally satisfies the criteria for $|\psi_{\text{ini}}\rangle$ with $2\gamma = \gamma_H$. These observations verify (i) and (ii) in Problem~\ref{prob:bqp2}.
Finally, since an eigenvalue $\lambda_{j,H}$ of $H$ is mapped to an eigenvalue $\lambda_{j,A}$ of $A$ according to the relation
$
\lambda_{j,A} = {i}{C^{-1}} (1 - f(\lambda_{j,H}))$, {we have the following relation}:
\begin{align}
\lambda_{\min,H}\leqslant a_H\Longleftrightarrow g_{\max}\geqslant4/3C,\\
\quad \lambda_{\min,H}\geqslant b_H\Longleftrightarrow g_{\max}\leqslant2/3C.
\end{align}
Thus, (iii) in Problem.~\ref{prob:bqp2} is satisfied with $\varepsilon^{-1} = {3C}/{2} = O(\mathrm{poly}(n))$. 
In conclusion, each instance of the GHL problem can be efficiently reduced to an instance of Problem~\ref{prob:bqp2}, and therefore, the classical hardness of the GHL problem implies the classical hardness of Problem~\ref{prob:bqp2}.

Now, we show that Problem~\ref{prob:bqp2} can be efficiently solved on a quantum computer. The key challenge is that we are only given an initial state $|\psi'_{\text{ini}}\rangle$ with nontrivial overlap to certain eigenvectors of $A$, which is a weaker condition compared to that required in FQED, where $|\psi_{\text{ini}}\rangle$ is assumed to have nontrivial overlap to the low singular-value subspace. Fortunately, we can establish the following result, which relates these two conditions (see the proof in the End Matter).

\begin{lemma}\label{lm:ul}
Let $(\lambda, |v\rangle)$ be an arbitrary eigenvalue-eigenvector pair of $A$. For an arbitrary quantum state $|\psi\rangle$, if $|\langle v|\psi\rangle| \geqslant 2\gamma$ and $|\mu - \lambda| \leqslant \varepsilon \gamma$, then
$
\left\| \Pi^{(\tilde{A})}_{\varepsilon} |\psi\rangle \right\| \geqslant \gamma$.
\end{lemma}

\noindent By definition, $|\psi'_{\text{ini}}\rangle$ has nontrivial overlap with an eigenvector $|v\rangle \in \mathcal{V}_{\text{complex}}$. If we are further promised that $\mu$ is sufficiently close to the corresponding eigenvalue (i.e., $|\mu - \lambda| \leqslant \varepsilon \gamma$), then the condition in Eq.~\eqref{eq:overlap} required for FQED is satisfied. 
Therefore, we can use $|\psi'_{\text{ini}}\rangle$ as the input state, and design a set of FQED instances such that their inner disks cover the region corresponding to the complex spectrum, while their outer disks do not overlap with the region corresponding to the real spectrum. By determining whether there exists any FQED instance that returns \texttt{True}, we can efficiently distinguish between the real and complex spectrum cases.
Finally, since the GLH problem has been proven to be BQP-hard~\cite{Gharibian.22}, we thus establish the BQP-completeness of Problem~\ref{prob:bqp2}.

\begin{figure} 
	\centering	\includegraphics[width=.5\textwidth]{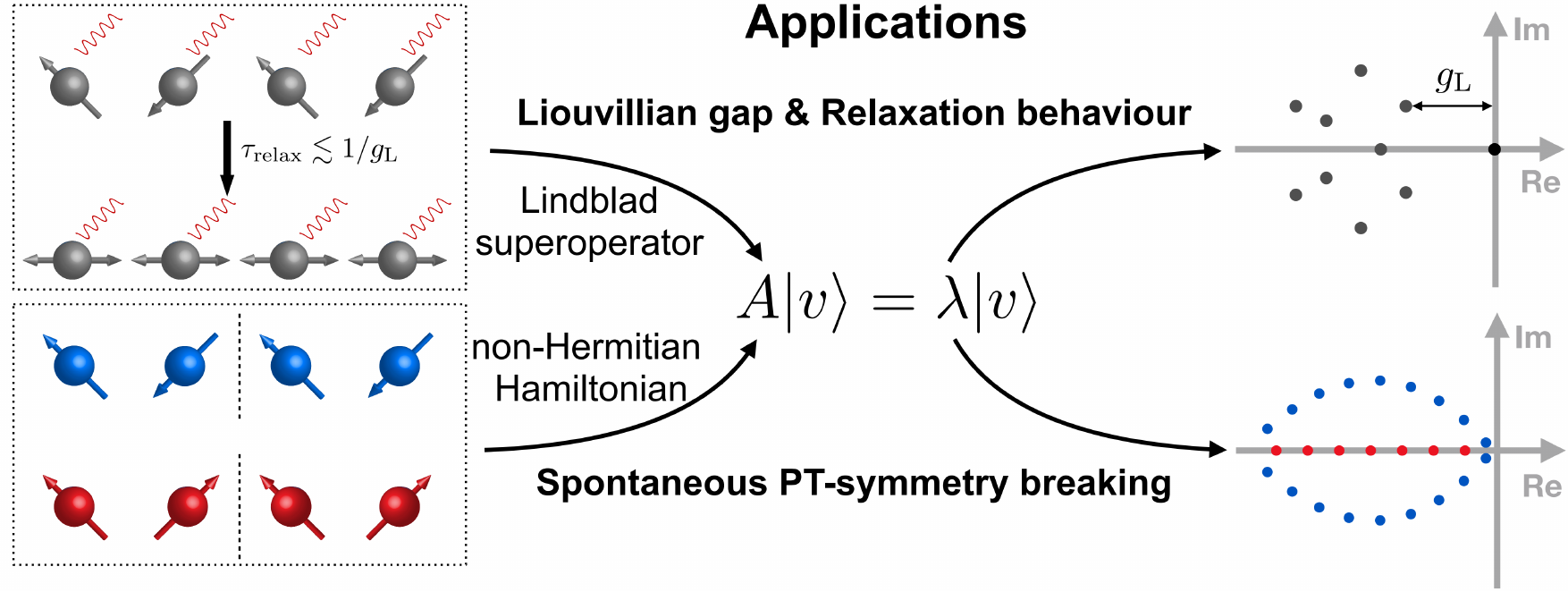} 
	\caption{{Applications of eigenvalue problems in non-Hermitian many-body physics.}}
	\label{fig:sketch} 
\end{figure}

\vspace{.2cm}
\noindent \textbf{\textit{Applications.}}
Since non-Hermitian Hamiltonians describe a wide range of physical and classical phenomena, our algorithm has broad applications. {Here, we introduce three representative examples and leave detailed discussions to~\cite{sm}.}

First, the Liouvillian gap is critical for understanding the relaxation time of an open quantum system~\cite{Medvedyeva.16,Banchi.17,Rowlands.18,Mori.20,Yuan.21,Zhou.22} and the efficiency of quantum algorithms based on dissipation~\cite{ding2024single,lin2025dissipative,zhan2025rapid}. Let $A_L$ denote the vectorized Liouvillian superoperator of a master equation. The Liouvillian gap $g_L$ is defined as the smallest nonzero distance from an eigenvalue of $A_L$ to the real axis. This quantity is of practical importance, as the relaxation time $\tau$ is typically characterized by $\tau \lesssim 1 / g_L$~\cite{Mori.20}. However, analytic solutions for $g_L$ exist only in certain special cases, and classical numerical computation becomes intractable due to the exponential scaling of the Hilbert space. In contrast, estimating $g_L$ can be formulated as a line gap problem, which can be efficiently addressed using our method.

Second, non-Hermitian Hamiltonians can often be viewed as an approximation to the full Lindbladian master equation~\cite{Dalibard.92, Carmichael.93,Nakagawa.18,Song.19}, typically by neglecting the quantum jump operators. It provides an important framework for studying topological, symmetry, and other key properties. As a notable example, a parity-time ($PT$) symmetric Hamiltonian $A$ may be in either the $PT$-broken or $PT$-unbroken phase. The distinction between the $PT$-broken and $PT$-unbroken phases corresponds directly to the complex versus real spectrum cases in Problem~\ref{prob:bqp2}~\cite{Bender.98,Bender.99,Khare.00,Delabaere.00,Mostafazadeh.02,Mostafazadeh.02_}. %Therefore, our algorithm can potentially provide an exponential speedup in characterizing $PT$-symmetry properties of many-body open quantum systems.

Third, Markov chains have widespread applications across natural and social sciences~\cite{Norris.98,Odencrantz.00,Ibe.12,Meyn.12,Levin.17}, which can be represented by a non-Hermitian stochastic matrix $A$, whose largest eigenvalue is $1$. The absolute spectral gap of $A$ is defined as $g_{\text{ag}} \equiv 1 - \max_{\lambda_j \neq 1} |\lambda_j|$, which determines the relaxation time $t_{\text{ag}} \equiv 1 / g_{\text{ag}}$ of the corresponding stochastic process. In particular, for an ergodic, time-reversible Markov chain (i.e., one satisfying the detailed balance condition), $t_{\text{ag}}$ serves as an upper bound for the mixing time, namely the time required to converge to the stationary distribution~\cite{Levin.17}. By applying a strategy similar to that used for solving Problem~\ref{prob:lg}, we can efficiently estimate $g_{\text{ag}}$ and hence characterize the relaxation behavior of such systems.

\vspace{.2cm}
\noindent \textbf{\textit{Discussions.}} We have introduced an efficient quantum algorithm for solving non-Hermitian eigenproblems and demonstrated their classical computational hardness. To further illustrate the generality of our approach, we provide additional discussions in~\cite{sm}, including the following directions:
(i) Generalization to nondiagonalizable matrices, as well as simplifications in certain cases, such as when the eigenvalues are promised to be real, {and} when only an arbitrary eigenvalue is required.
(ii) Quantum algorithms for point gap problem, i.e., finding the eigenvalue closest to a given point, using a similar strategy and achieving comparable runtime.
(iii) Efficient implementations of the applications introduced above. 

\vspace{0.2cm}
\noindent \textbf{\textit{Acknowledgement.}} We thank Seth Lloyd, Xiaogang Li and Dong Yuan for their helpful discussions. X.Y and Y.Z are supported by the Innovation Program for Quantum Science and Technology (Grant No.~2023ZD0300200), the National Natural Science Foundation of China (Grants No.~12175003, No.~12361161602, and No.~12247124), NSAF (Grant No.~U2330201). X.-M.Z is supported by National Natural Science Foundation of China (Grant No.~12405013) and Project funded by China Postdoctoral Science Foundation (Grant No. 2023T160004). The work is supported by the High-performance Computing Platform of Peking University.

%This work is supported by the Innovation Program for Quantum Science and Technology (Grant No.~2023ZD0300200), the National Natural Science Foundation of China (Grant No.~12405013, No.~12175003, No.~12361161602, and No.~12247124), NSAF (Grant No.~U2330201), and Project funded by China Postdoctoral Science Foundation (Grant No. 2023T160004).

\vspace{0.2cm}
\noindent \textit{Note-added.} Another result, Theorem.~12 in Ref~\cite{Low.24}, has appeared concurrently to solve an eigenproblem similar to Problem~\ref{prob:lg}, based on a general framework to realize analytical functions of $A$. Their result relies on stronger assumptions, that initial state is at most $O(\varepsilon)$ distance to the corresponding eigenvector, and the absolute value of the target eigenvalue is known a prior. 

%\bibliography{refs_eg}

\newpage

\onecolumngrid

\begin{centering}

\textbf{End Matter}

\end{centering}

$\\$

\twocolumngrid

\noindent\textbf{\textit{Proof of Eq.~\eqref{eq:cu}}.}
$C(\mu)\leqslant\min_{\lambda_j}|\mu-\lambda_j|$ follows from Weyl's Theorem~\cite{Horn.94}. So we focus on the upper bound of $\min|\mu-\lambda_j|$. 

When $\mu=\lambda_j$, Eq.~\eqref{eq:cu} holds obviously. When $\mu\neq\lambda_j$, the cost function is equivalent to
$C(\mu)=\left\|(A-\mu I)^{-1}\right\|^{-1}$. From Jordan decomposition, we have 
$(A-\mu I)^{-1}=P(\Lambda-\mu I)^{-1} P^{-1}$, where $\Lambda=\text{diag}(\lambda_{0},\lambda_{1},\cdots)$. So
\begin{align}
\|(A-\mu I)^{-1}\|&\leqslant \|P\|\|(\Lambda-\mu I)^{-1}\| \|P^{-1}\|\notag\\
&\leqslant K\|(\Lambda-\mu I)^{-1}\|. \label{eq:k1}
\end{align}
The last inequality is due to $\kappa\equiv\|P\|\|P^{-1}\|\leqslant K$. Combining with 
 \begin{align}\label{eq:k3}
 \left\|(\Lambda-\mu I)^{-1}\right\|=\frac{1}{\min|\mu-\lambda_j|},
 \end{align}
we obtain $\min_{\lambda_j}|\mu-\lambda_j|\leqslant K C(\mu)$.

%\section{Proof of Lemma~\ref{lm:ul}}

\noindent\textbf{\textit{Proof of Lemma~\ref{lm:ul}.}}
We decompose $|v\rangle$ and $|\psi\rangle$ using the right singular vector of $\tilde{A}\equiv A-\mu I$ as follows
\begin{align}
|v\rangle=\sum_{j}\alpha_{j}(v)|u_j\rangle,\quad\quad
|\psi\rangle=\sum_{j}\alpha_{j}(\psi)|u_j\rangle,
\end{align}
which by definition satisfies $\left|\sum_{j}\alpha_j(v)\alpha_j(\psi)\right|\geqslant2\gamma$. According to the triangular inequality, we also have
\begin{align}\label{eq:gam0}
&\sum_{j}\left|\alpha_j(v)\alpha_j(\psi)\right|\geqslant\big|\sum_{j}\alpha_j(v)\alpha_j(\psi)\big|\geqslant2\gamma.
\end{align}
Because $|\mu-\lambda|\leqslant\varepsilon\gamma$, we have $\left\|(A-\mu I)|v\rangle\right\|\leqslant\varepsilon\gamma$, which is equivalent to $\sqrt{\sum_{j}\tilde{\sigma}_j^2|\alpha_j(v)|^2}\leqslant\varepsilon\gamma$, and hence  $\varepsilon^{-1}\sqrt{\sum_{j}\tilde{\sigma}_j^2|\alpha_j(v)|^2}\leqslant\gamma$. Accordingly, we have
\begin{align}
&\sqrt{\sum_{j\in\{j':\tilde{\sigma}_{j'}>\varepsilon\}}|\alpha_j(v)|^2}\notag\\
\leqslant&\varepsilon^{-1}\sqrt{\sum_{j\in\{j':\tilde{\sigma}_{j'}>\varepsilon\}}\tilde{\sigma}_j^2|\alpha_j(v)|^2}\notag\\
\leqslant&\gamma.
\end{align}
Using Cauchy-Schwarz inequality and notice that $\sqrt{\sum_{j\in\{j':\tilde{\sigma}_{j'}>\varepsilon\}}|\alpha_j(\psi)|^2}\leqslant1$, we have
\begin{align}
&\sum_{j\in\{j':\tilde{\sigma}_{j'}>\varepsilon\}}|\alpha_j(v)\alpha_j(\psi)|\notag\\
\leqslant&\sqrt{\sum_{j\in\{j':\tilde{\sigma}_{j'}>\varepsilon\}}|\alpha_j(v)|^2}\sqrt{\sum_{j\in\{j':\tilde{\sigma}_{j'}>\varepsilon\}}|\alpha_j(\psi)|^2} \notag\\
\leqslant&\gamma.
\end{align}
Combining with Eq.~\eqref{eq:gam0}, we have 
\begin{align}\label{eq:gam1}
\sum_{j\in\{j':\tilde{\sigma}_{j'}\leqslant\varepsilon\}}\left|\alpha_j(v)\alpha_j(\psi)\right|&\geqslant2\gamma-\sum_{j\in\{j':\tilde{\sigma}_{j'}>\varepsilon\}}\left|\alpha_j(v)\alpha_j(\psi)\right|\notag\\
&\geqslant\gamma.
\end{align}
Using Cauchy-Schwarz inequality again, we have

\begin{align}\label{eq:gam2}
&\sum_{j\in\{j':\tilde{\sigma}_{j'}\leqslant\varepsilon\}}\left|\alpha_j(v)\alpha_j(\psi)\right|\notag\\
\leqslant&\sqrt{\left(\sum_{j\in\{j':\tilde{\sigma}_{j'}\leqslant\varepsilon\}}\left|\alpha_j(v)\right|^2\right)\left(\sum_{j\in\{j':\tilde{\sigma}_{j'}\leqslant\varepsilon\}}\left|\alpha_j(\psi)\right|^2\right)}\notag\\
\leqslant&\sqrt{\left(\sum_{j\in\{j':\tilde{\sigma}_{j'}\leqslant\varepsilon\}}\left|\alpha_j(\psi)\right|^2\right)}\notag\\
=& \left\|\Pi^{(\tilde{A})}_{\varepsilon}|\psi\rangle\right\|.
\end{align}
Combining Eq.~\eqref{eq:gam1} with Eq.~\eqref{eq:gam2}, we have
\begin{align}
\left\|\Pi^{(\tilde{A})}_{\varepsilon}|\psi\rangle\right\|\geqslant\gamma.
\end{align}

%\section{QSVT for singular value threshold problem}
\noindent\textbf{\textit{QSVT for singular value threshold problem.}}
To begin with, we show that given the block encoding of $A$, (i.e. $\mathscr{O}_A$), we are able to construct the block encoding of $\tilde{A}=A-\mu I$, up to a small normalization factor. We introduce one extra ancillary qubit, and define $R_{\text{anc}}$ as single qubit rotation $\begin{pmatrix}\cos\theta&-\sin\theta\\
\sin\theta&\cos\theta
\end{pmatrix}$ applied at it with
 $\theta=\arccos\left((1+|\mu|)^{-1/2}\right)$. Let
\begin{align}\label{eq:oamu}
\mathscr{O}_{\tilde{A}}=R^{\dag}_{{\rm anc}}\left(|0\rangle\langle0|\otimes \mathscr{O}_{A}-|1\rangle\langle1|\otimes e^{i\text{arg}(\mu)}I\right)R_{{\rm anc}},
\end{align}
it can be verified that Eq.~\eqref{eq:oamu} is a block encoding of matrix $\tilde{A}/(1+|\mu|)$ and can be constructed with gate count $\mathcal{O}(C_{\rm be})$. 
Then, we show how to solve SVT problem based on $\mathscr{O}_{\tilde{A}}$, which mainly follows~\cite{Gilyen.19}.

Let $P=|0^{n_{\text{anc}}}\rangle\langle0^{n_{\text{anc}}}|\otimes I$ be the projection of the ancillary system to state $|0^{n_{\text{anc}}}\rangle$, we further define $R_{\phi}\equiv e^{i\phi(2P-I_{\text{anc}})}$ as the partial reflection operation along state $|0^{n_{\text{anc}}}\rangle$, where $I_{\text{anc}}$ is identity in the ancillary system. For some odd $K$, we let $\bm{\phi}=[\phi_{1},\phi_2,\cdots,\phi_K]$ be a set of angles. QSVT represents the following quantum circuit 
\begin{align}\label{eq:qsvt}
U_{\bm{\phi}}=R_{\phi_K}\mathcal{O}_{\tilde{A}}\prod_{k=1}^{(K-1)/2}R_{\phi_k}\mathcal{O}_{\tilde{A}}^{\dag}R_{\phi_k}\mathcal{O}_{\tilde{A}}.
\end{align}
Suppose the singular-value decomposition of $\tilde{A}$ is $\tilde{A}=\sum_{j}\tilde{\sigma}_j|u_j\rangle\langle v_j|$, it can be verified that $U_{\bm{\phi}}$ is in the form of 

\begin{align}\label{eq:main_poly}
U_{\bm{\phi}}= \begin{pmatrix}\text{Poly}_{\bm{\phi}}(\tilde{A})/(1+|\mu|)&*\\
*&*\end{pmatrix},
\end{align}
where
$\text{Poly}_{\bm{\phi}}(\tilde{A})\equiv\left(\sum_{j}\text{Poly}_{\bm{\phi}}(\tilde{\sigma}_j)|u_j\rangle\langle v_j|\right)$ for some $K$-degree polynomial function $\text{Poly}_{\bm{\phi}}(\cdot)$.
{In our SVT problem, the goal is to distinguish whether there is a small enough singular-value or not. To this end, we can construct a polynomial function filtering small singular values, with the following established result.}

\begin{lemma}[Adapted from Theorem 19 in~\cite{Gilyen.19}]\label{lm:filter}
For arbitrary $\varepsilon_{\text{th}}\in(0,0.5)$, there exists a set of angles $\bm{\phi}_{{\rm F}}=[\phi_1,\phi_2,\cdots,\phi_K]$ for some odd $K=\mathcal{O}(\varepsilon_{{\rm th}}^{-1}\log(\delta^{-1}))$, such that
\begin{eqnarray}\label{eq:sign}
{\rm Poly}_{\bm{\phi}_{{\rm F}}}(x)\in \left\{
\begin{array}{lcl}
\ [1-\delta,1]      &    &  x\leq\varepsilon_{{\rm th}}\\
\ [0,1]     &    &  \varepsilon_{{\rm th}}<\sigma_0\leq2\varepsilon_{{\rm th}}\\
\ [0,\delta]   &  &x>2\varepsilon_{{\rm th}}
\end{array} \right.
\end{eqnarray}
\end{lemma}

\noindent {Eq.~\eqref{eq:sign} can be considered as an approximation of the shifted Heaviside function. }
The explicit construction of $\bm{\phi}_K$ is referred to~\cite{low2017}. For simplicity, we denote the corresponding QSVT as $U_{\rm F}\equiv U_{\bm{\phi}_{{\rm F}}}$.
Recall that the initial state $|\psi_{\text{ini}}\rangle$ has a nontrivial overlap to the low singular value subspace (if it exists) as represented in Eq.~\eqref{eq:overlap}.
By applying $U_{\text{F}}$ to $|0^{n_{\text{anc}}}\rangle|\psi_{\text{ini}}\rangle$, we obtain
\begin{eqnarray}\label{eq:out}
U_{\text{F}}|0\rangle|\psi_{\rm{ini}}\rangle\approx \left\{
\begin{array}{lcl}
\gamma'|0^a\rangle|\psi'\rangle+|\rm{garb}\rangle,      &    &  C(\mu)\leqslant\varepsilon_{{\rm th}}\\
|\rm{garb}'\rangle,    &  &C(\mu)\geqslant2\varepsilon_{{\rm th}}
\end{array} \right.
\end{eqnarray}
for some $\gamma'\geqslant\gamma/(1+|\mu|)\geqslant\gamma/2$. Here, the short hand $|\psi_{a}\rangle\approx|\psi_{b}\rangle$ means that $|\langle\psi_{a}|\psi_{b}\rangle|\leqslant1-\delta$, and the unnormalized garbage states satisfy $P|\text{garb}\rangle=P|\text{garb}'\rangle=0$, where $P=|0^a\rangle\langle0^a|\otimes I$. The \texttt{True} and \texttt{False} output correspond to the first and second lines of Eq.~\eqref{eq:out}, and their amplitudes of $|0^a\rangle|\psi'\rangle$ are either $\geqslant\gamma(1-\delta)/2$ or $\leqslant\delta$. Note that $\delta$ is a sufficiently small constant. Using binary amplitude estimation technique (Lemma.12 in~\cite{Dong.22}), one can distinguish these two cases with totally $\tilde{\mathcal{O}}(\gamma^{-1})$ implementations of $U_F$ and unitary $\mathscr{P}_{\tilde{A}}$ that preparing $|\psi_{\text{ini}}\rangle$.

Each $U_F$ contains $\tilde{\mathcal{O}}(\varepsilon_{\text{th}}^{-1})$ queries to $\mathscr{O}_{\tilde{A}}$ and $R_{\phi_k}$. The latter contains $\mathcal{O}(n)$ number of extra single- and  two-qubit gates. So $U_{\text{F}}$ has gate count $\tilde{\mathcal{O}}(\varepsilon^{-1}_{\text{th}}(C_{\text{be}}+n))$. Moreover, the gate count of state preparation $\mathscr{P}_{\tilde{A}}$ is $C_{\text{sp}}$. So the total gate count of solving SVT problem (thus constructing FQED subroutine) is $\tilde{\mathcal{O}}(\gamma^{-1})\times\big(\tilde{\mathcal{O}}(\varepsilon^{-1}_{\text{th}}(C_{\text{be}}+n))+C_{\text{sp}}\big)=\tilde{\mathcal{O}}(\gamma^{-1}(\varepsilon_{\text{th}}^{-1}(C_{\text{be}}+n)+C_{\text{sp}}))$, as claimed in Eq.~\eqref{eq:gc0}.

\newpage

\setcounter{secnumdepth}{3}  %enforce numbering of section in PRL template, we keep PRL template because we like the way it labels references
\setcounter{equation}{0}%reset counter
\setcounter{figure}{0}
\setcounter{table}{0}
\setcounter{section}{0}

\renewcommand{\theequation}{S-\arabic{equation}}
\renewcommand{\thefigure}{S\arabic{figure}}
\renewcommand{\thetable}{S-\Roman{table}}
\renewcommand\figurename{Supplementary Figure}
\renewcommand\tablename{Supplementary Table}
\newcommand\citetwo[2]{[S\citealp{#1}, S\citealp{#2}]}
\newcommand\citecite[2]{[\citealp{#1}, S\citealp{#2}]}

\newcolumntype{M}[1]{>{\centering\arraybackslash}m{#1}}
\newcolumntype{N}{@{}m{0pt}@{}}

\makeatletter \renewcommand\@biblabel[1]{[S#1]} \makeatother

\makeatletter \renewcommand\@biblabel[1]{[S#1]} \makeatother

%%%%%%%%%%%%%%%%%%%%%%%%%%%%%%%%%%
% The supplementary text starts here
%%%%%%%%%%%%%%%%%%%%%%%%%%%%%%%%%%
\onecolumngrid
\begin{center}
{\bf\large Supplemental material}
\end{center}
%\vspace{0.5cm}

\tableofcontents

%\newpage
%\begin{appendix}

\setcounter{lemma}{0}

\section{Fuzzy quantum eigenvalue detector generalized for nondiagonalizable matrices}

 Given a general matrix $A$, we consider its Jordan canonical form  $A=P\Lambda P^{-1}$, where $\Lambda\equiv\Lambda\equiv\Lambda_1\oplus\Lambda_2\oplus\cdots\oplus\Lambda_M$ is a block-diagonal matrix, whose Jordan blocks are  in the form of
\begin{align}
\Lambda_j=
\begin{pmatrix}
\lambda_j&1\\
&\lambda_j&\ddots\\
&&\ddots&1\\
&&&\lambda_j
\end{pmatrix}.
\end{align}
We say that $A$ is diagonalizable if $\max(\text{dim}(\Lambda_j))=1$. If there exist a Jordan block with dimension larger than $1$, we say that $A$ is non-diagonalizable, or defective. We assume that we are promised that $\max(\text{dim}(\Lambda_j))\leqslant m_{\max}$ for some integer $m_{\max}$, which characterizes the defectiveness of $A$.

Recall that $C(\mu)$ is the minimum singular value of matrix $A-\mu I$, the following lemma generalizes Eq.~\eqref{eq:cu} in the main text.

\begin{lemma}\label{lm:main1}
For an arbitrary matrix $A$, we have 
\begin{align}\label{eq:cus}
C(\mu)\leqslant\min_{\lambda_j}|\mu-\lambda_j|\leqslant 3(KC(\mu))^{1/m_{\max}}.
\end{align}
\end{lemma}

\begin{proof}
Again, $C(\mu)\leqslant\min_{\lambda_j}|\mu-\lambda_j|$ follows directly from Weyl's theorem, so we only consider the second inequality.  When $C(\mu)=0$, Eq.~\eqref{eq:cus} holds obviously. For $C(\mu)>0$, we have
\begin{align}
C(\mu)=&\left\|(A-\mu I)^{-1}\right\|^{-1}\notag\\
=&\left\|P\text{diag}\left(\tilde\Lambda_1^{-1},\tilde\Lambda_2^{-1},\cdots,\tilde\Lambda_M^{-1}\right)P^{-1}\right\|^{-1}\notag\\
\geqslant&\kappa_P^{-1}\left\|\text{diag}\left(\tilde\Lambda_1^{-1},\tilde\Lambda_2^{-1},\cdots,\tilde\Lambda_M^{-1}\right)\right\|^{-1}\notag\\
\geqslant&\kappa_P^{-1}\left\|\tilde\Lambda_j^{-1}\right\|^{-1}\notag\\
=&\frac{\sigma_{\text{min}}(\tilde\Lambda_j)}{\kappa}.\label{eq:des0}
\end{align}
Note that Eq.~\eqref{eq:des0} is applied for arbitrary $j$. According to Ref.\cite{Kahan.82} (see also Ref.~\cite{Jiang.94}), let $\delta_j=|\lambda_j-\mu|$, we have

\begin{align}\label{eq:des1}
\sigma_{\min}(\tilde{\Lambda}_j)\geqslant\frac{\delta_j^{m_j}}{(1+\delta_j)^{m_j-1}},
\end{align}
where $m_j$ is the dimension of $\tilde{\Lambda}_j$. Because the operator norm of $A$ is bounded by $\|A\|\leqslant1$, we also have $|\lambda_j|\leqslant1$ for all eigenvalues. Our searching region is also restricted by $|\mu|\leqslant1$, so we have $\delta_j\leqslant2$. We can simplify Eq.~\eqref{eq:des1} as
\begin{align}\label{eq:des2}
\sigma_{\min}(\tilde{\Lambda}_j)\geqslant\left(\frac{\delta_j}{1+\delta_j}\right)^{m_j}(1+\delta_j)\geqslant(\delta_j/3)^{m_j}.
\end{align}
Combining Eq.~\eqref{eq:des0} with Eq.~\eqref{eq:des2}, we have
\begin{align}
\kappa C(\mu)\geqslant(\delta_j/3)^{m_j},
\end{align}
which gives
\begin{align}
\delta_j \leqslant  3(\kappa C(\mu))^{1/m_j}.
\end{align}
Because $\min_j|\mu-\lambda_j|\leqslant\delta_j$, we have
\begin{align}\label{eq:des3}
\min_j|\mu-\lambda_j| \leqslant  3(\kappa C(\mu))^{1/m_j}.
\end{align}
When $C(\mu)\leqslant 1/\kappa$, we have $\kappa_PC(\mu)\leqslant1$, and the right hand side of Eq.~\eqref{eq:des3} increases monotonically with $m_j$. So $\min_j|\mu-\lambda_j| \leqslant  3(\kappa C(\mu))^{1/m_{\max}}$. When $C(\mu)> 1/\kappa$, we have $\kappa C(\mu)>1$, so the right hand side of Eq.~\eqref{eq:des3} is larger than 3. Because we always have $\min_j|\mu-\lambda_j|\leqslant2$, so we also have $\min_j|\mu-\lambda_j| \leqslant  3(\kappa C(\mu))^{1/m_{\max}}$. Then, Lemma.~\ref{lm:main1} follows from $\kappa\leqslant K$.

\end{proof}

\noindent With the same argument in the main text, we are able to generalize the definition of Fuzzy quantum eigenvalue detector (FQED) to the following.

\vspace{.2cm}
 \begin{definition}[fuzzy quantum eigenvalue detector for general matrices]\label{def:fqedd}
 Let
\begin{eqnarray}\label{eq:sign}
\nu(r)=\left\{
\begin{array}{lcl}
r/(2K)        &  m_{{\rm max}}=1\\
\left(r/3\right)^{m_{\max}}(2K)^{-1} & m_{{\rm max}}>1
\end{array} \right.
\end{eqnarray}
 If $\min_j|\mu-\lambda_j|< \nu(r)$, $ { \rm FQED}(\mu,r)$ returns {\rm \texttt{True}}; if \, ${\rm FQED}(\mu,\nu(r))$ returns ` {\rm \texttt{True}}, we have $\min_j|\mu-\lambda_j|< r$.

\end{definition}

\vspace{.2cm}
\noindent Based on Lemma.~\ref{lm:main1}, we can construct FQED defined above with the same process introduced in the main text with the same runtime. 
\vspace{.2cm}

 \begin{lemma}\label{lm:FQED_}
For $|\mu|<1$ and $\varepsilon_{\rm th}\in(0,1)$, the subroutine ${\rm FQED}(\mu,\varepsilon_{\rm th})$ can be constructed with gate count
 \begin{align}\label{eq:cfqed}
 C_{{\rm FQED}}=\tilde{\mathcal{O}}\left(\gamma^{-1}\left(\varepsilon_{{\rm th}}^{-1}(C_{{\rm be}}+n)+C_{{\rm sp}}\right)\right)
 \end{align}
 with arbitrarily small failure probability $\delta$.
 \end{lemma}

\vspace{.2cm}
\noindent Here and after, whenever FQED is queried, we assume that the failure probability $\delta$ is set as a sufficiently small value, such that the failure case is negligible. This will always cost at most $\mathcal{O}(\text{polylog}(\delta^{-1}))$ complexity, which is absorbed in the expression $\tilde{\mathcal{O}}$. In the next section, we consider the point gap problem for general matrices that are not necessarily to be diagonalizable.  The treatment for line gap problem (i.e. Problem.~\ref{prob:lg} in main text) with defective matrices is basically the same, which will not be repeated.

\section{Quantum algorithm for point gap Problem}\label{sec:spg}
We consider the point gap problem as follows.

\vspace{.2cm}
\begin{problem}\label{prob:pgf}
 Given a matrix $A\in\mathbb{C}^{N\times N}$ with spectral norm $\|A\|\leqslant1$, and a reference point $\boldsymbol{P}$. The goal is to output an eigenvalue $\lambda_{\min}$ closest to $\boldsymbol{P}$ up to an accuracy $\varepsilon\in(0,1)$, with promised that the distance from $\lambda_{\min}$ to $\boldsymbol{P}$ is larger than $\varepsilon$.
\end{problem}
\vspace{.2cm}

Similar to the argument in the main text for line gap problem, we can specify the reference point as the original point, i.e. $\bm{P}=0$. Solutions to other reference points are fundamentally the same.  Initially, we set the guess region of the gap as $g'\in[\varepsilon,1+\varepsilon]$. 
 We introduce an eigenvalue range shrinking subroutine (ERSS) in Sec.~\ref{sec:erss}, based on which the guess region of $g$ shrinks iteratively. More specifically, we first set $R^{\min}_0=\varepsilon$ and $R^{\max}_0=1+\varepsilon$. Suppose that at the $j$th step, we have
\begin{align}
g'\in[R^{\min}_{j-1},R^{\max}_{j-1}].
\end{align}
In this step, $R^{\min}_{j-1}$ or $R^{\max}_{j-1}$ is updated by querying ERSS
\begin{align}
\mathscr{S}_{\text{ring}}(R^{\min},R^{\max},r)&\rightarrow(\tilde{R}^{\min},\tilde{R}^{\max})
\end{align}
 defined in Algorithm.~\ref{alg:erss0} (see also Sec.~\ref{sec:erss} below). The ERSS contains three input parameters. $R^{\min}$ and $R^{\max}$ characterizes the recent confidence region in which the value of gap is in. The third parameter $r>0$ controls the step size of updating. It is required that $r\leqslant R^{\min}$ and $r\leqslant R^{\max}-R^{\min}$. The first requirement ensures that the output of ERSS will not be affected by eigenvalue at the original point, if any. The second requirement ensures that the  gap $\Delta_j\equiv R^{\max}_j-R^{\min}_j$ reduces monotonically with $j$. 
 The ERSS has the following property.
\newline 

\begin{lemma}\label{lm:s2}
Let $A$ be a square matrix satisfying $\|A\|\leqslant1$. Let $\left(\tilde R_a, \tilde R_b\right)$ be the output of $\mathscr{S}_{\text{ring}}(R_{a},R_b,r)$ for some $0<R_a<R_b\leqslant1$, and $0<r\leqslant\min(R_a,(R_b-R_a)/2)$. Then, suppose the point gap satisfies $g\in[R_a,R_b]$, we have $g\in[\tilde R_a, \tilde{R}_b]$. Here, $\tilde R_a$ and $\tilde{R}_b$ are defined in Algorithm.~\ref{alg:erss0}.
\end{lemma} 

$\\$
\noindent Moreover, the complexity of ERSS is given by the following. \newline

\begin{lemma}\label{lm:rg_rt}
$\mathscr{S}_{\text{ring}}(R_{a}, R_{b},r)$ defined in Algorithm.~\ref{alg:erss0} can be realized with quantum gate count 

\begin{align}\label{eq:cerss}
C_{\rm ERSS}=\tilde{\mathcal{O}}\left(R_{a} K\gamma^{-1}r^{-m_{\text{max}}} (Kr^{-m_{\text{max}}}(C_{\rm be}+n)+C_{\rm sp}\right).
\end{align}
% queries to $\mathscr{O}_A$, $\mathscr{P}_A$ and there inverses, and single- and two-qubit gates.
\end{lemma}
$\\$

\noindent Base on Lemma~\ref{lm:s2}, we update guess region of $g'$ by $\mathscr{S}_{\text{ring}}\left(R_{j-1}^{\min},R_{j-1}^{\max},r\right)\rightarrow(R_j^{\min},R_j^{\max})$. %, where $\delta'$ is set as a sufficiently small value. 
From Algorithm.~\ref{alg:erss0}, it can also be verified that 
\begin{align}\label{eq:delta_c}
|\Delta_{j-1}-\Delta_{j}|=\Omega(r^{m_{\max}}/K).
\end{align}

Our algorithm is separated into two stages. In stage 1, we set $r=R_{j-1}^{\min}$ at each step, and this stage terminates when $R_{j}^{\min}\geqslant R_{j}^{\max}/2$. Using $R^{\min}_j\geqslant\varepsilon$, the complexity of each step is upper bounded by 
$
\tilde{\mathcal{O}}\left(K\gamma^{-1}\varepsilon^{-m_{\text{max}}+1} (K\varepsilon^{-m_{\text{max}}}(C_{\rm be}+n)+C_{\rm sp})\right).
$
From Eq.~\eqref{eq:delta_c}, it can be verified that this stage terminates with at most $O(K\varepsilon^{-m_{\max}+1})$ steps. So the total complexity for this substage is 
$\tilde{\mathcal{O}}\left(K^2\gamma^{-1}\varepsilon^{-2m_{\text{max}}+2} (K\varepsilon^{-m_{\text{max}}}(C_{\rm be}+n)+C_{\rm sp})\right).$

In stage 2, we set $r=\left(R_{j-1}^{\max}-R_{j-1}^{\min}\right)/2$, and this stage terminates when $\Delta_j\leqslant\varepsilon$. The final output, estimated minimum eigenvalue, is just the center of the last FQED with output True. In this stage, the complexity of each step is 
$
\tilde{\mathcal{O}}\left(K\gamma^{-1}\varepsilon^{-m_{\text{max}}} (K\varepsilon^{-m_{\text{max}}}(C_{\rm be}+n)+C_{\rm sp})\right).
$
 This stage contains $\tilde O(K\varepsilon^{-m_{\max}+1})$ steps as can be verified from Eq.~\eqref{eq:delta_c}. So the total complexity of stage 2 is
$\tilde{\mathcal{O}}\left(K^2\gamma^{-1}\varepsilon^{-2m_{\text{max}}+1} (K\varepsilon^{-m_{\text{max}}}(C_{\rm be}+n)+C_{\rm sp})\right)$.
Combining both stages, the total complexity of solving point gap problem is
\begin{align}\label{eq:opg}
\tilde{\mathcal{O}}\left(K^2\gamma^{-1}\varepsilon^{-2m_{\text{max}}+1} (K\varepsilon^{-m_{\text{max}}}(C_{\rm be}+n)+C_{\rm sp})\right).
\end{align}
For diagonalizable matrix, the complexity reduces to 
\begin{align}
\tilde{\mathcal{O}}\left(K^2\gamma^{-1}\varepsilon^{-1} (K\varepsilon^{-1}(C_{\rm be}+n)+C_{\rm sp})\right),\notag
\end{align}
which is identical to the complexity for diagonalizable line gap problem given in the main text. The process of generalizing line gap problems for non-diagonalizable matrices is similar to  the process above, whose gate complexity is also given by Eq.~\eqref{eq:opg}.

\begin{algorithm} [H]
\caption{ Algorithm for solving Problem~\ref{prob:pgf}}  
\label{alg:rg1}  
\begin{algorithmic}

\STATE $R_0^{\min}\leftarrow\varepsilon$; $R_0^{\max}\leftarrow1+\varepsilon$; $j\leftarrow1$

\STATE \textbf{while} $R_{j-1}^{\min}< R_{j-1}^{\max}/2$: $\quad\quad\quad\quad\quad\quad\quad\quad\quad\quad\quad\quad\quad\quad\quad\#$ stage 1

\STATE \quad $(R_{j}^{\min},R_{j}^{\max})\leftarrow\mathscr{S}_{\text{ring}}(R_{j-1}^{\min},R_{j-1}^{\max}, R_{j-1}^{\min})$

\STATE \quad $j\leftarrow j+1$

\STATE \textbf{end while} 

\STATE \textbf{while} $R_{j-1}^{\max}-R_{j-1}^{\min}>\varepsilon$: 
$\quad\quad\quad\quad\quad\quad\quad\quad\quad\quad\quad\quad\quad\quad\;\#$ stage 2

\STATE \quad $(R_{j}^{\min},R_{j}^{\max})\leftarrow\mathscr{S}_{\text{ring}}\left(R_{j-1}^{\min},R_{j-1}^{\max}, \left(R_{j-1}^{\max}-R_{j-1}^{\min}\right)/2\right)$

\STATE \quad $j\leftarrow j+1$

\STATE \textbf{end while}

\STATE \textbf{return} $\left(R_{j-1}^{\min},R_{j-1}^{\max}\right)$

\end{algorithmic} 
\end{algorithm} 

\subsubsection{Eigenvalue range shrinking subroutine}\label{sec:erss}
Here, we give detailed construction of the ERSS.

\begin{algorithm} [H]
\caption{ $\mathscr{S}_{\text{ring}}(R_{\text{a}},R_{\text{b}},r)$ (Eigenvalue range shrinking subrutine)}  
\label{alg:erss0}  
\begin{algorithmic}

%\STATE $\delta'\leftarrow\delta/|\mathcal{N}_{\text{ring}}(R_{\text{a}},\tilde{\nu}(r))|$
\STATE \textbf{for all} $t\in \mathcal{N}_{\text{ring}}(R_{\text{a}},\nu(r))$:

\STATE \quad $B\leftarrow \text{FQED}(t,\nu(r))$

\STATE \quad \textbf{if} $B=$ True:

\STATE \quad\quad \textbf{break for }

\STATE \quad \textbf{end if}

\STATE \textbf{end for} 

\STATE \textbf{if}  $B=$ True:

\STATE \quad  $\tilde R_{\text{a}}\leftarrow R_{\text{a}}$
\STATE \quad  $\tilde R_{\text{b}}\leftarrow R_{\text{a}}+r$

\STATE \textbf{else if}  $B=$ False:

\STATE \quad $\tilde R_{\text{a}}\leftarrow R_{\text{a}}+\nu(r)/2$
\STATE \quad  $\tilde R_{\text{b}}\leftarrow R_{\text{b}}$

\STATE \textbf{end if}

\STATE \textbf{return} $\left(\tilde R_{\text{a}}, \tilde R_{\text{b}}\right)$ 

\end{algorithmic} 
\end{algorithm}

\begin{figure}[t]
    \centering
          \includegraphics[width=1\columnwidth]{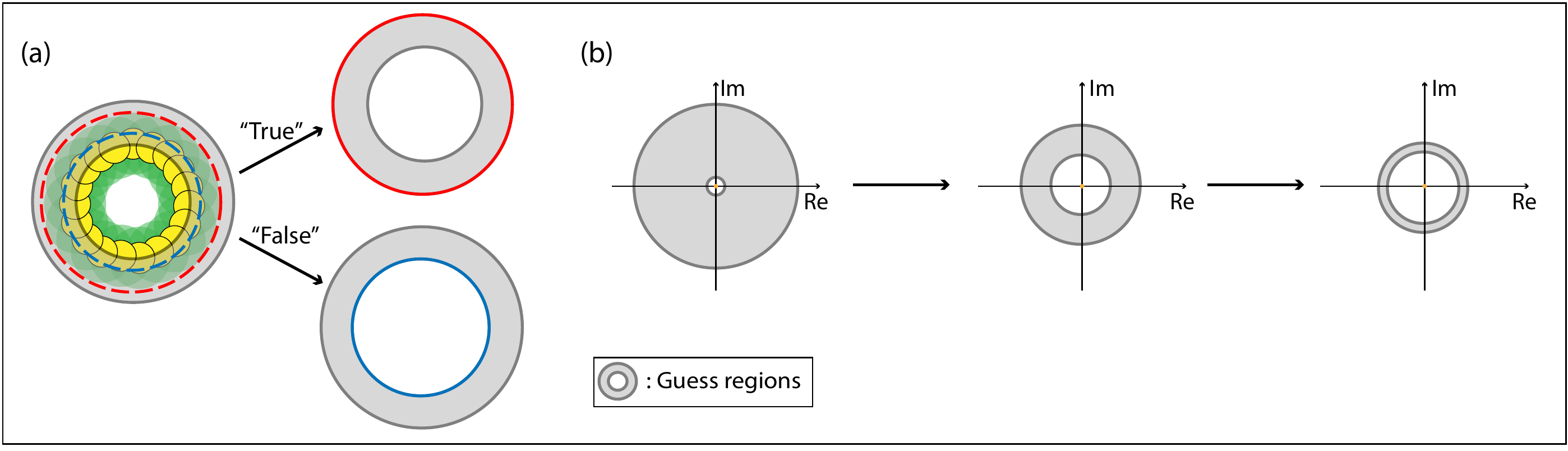}
       \caption{ Sketch of the process of solving Problem.~\ref{prob:pgf}. The initial guess region is a ring enclosed by two grey circles. At each iteration, we query a set of FQEDs which are encapsulated as $\mathscr{S}_{\text{ring}}$ in Algorithm.~\ref{alg:erss0}. If one of the FQEDs returns \texttt{True}, the ERSS returns \texttt{True}, otherwise the ERSS returns \texttt{False}. The updated guess region is a ring enclosed by a grey circle and red or blue circles. This process is repeated iteratively until the guess region is sufficiently small.} 
       \label{fig:pl}
\end{figure}

Here, we have also defined 
\begin{align}\label{eq:T1}
\mathcal{N}_{\text{ring}}(R,s)=\left\{Re^{i2\pi m/M(R,s)}\big|m\in\{1,2\cdots, M(R,s)\}\right\},
\end{align}
where 
\begin{align}\label{eq:T2}
M(R,s)=\frac{2\pi}{\text{arctan}(s/(2R))}.
\end{align}
$\mathcal{N}_{\text{ring}}(R,s)$ defines a set of points at the circle with radius $R$. The main idea of ERSS is illustrated in Fig.~\ref{fig:pl}(a). The inner (yellow) disks of all FQEDs cover the edge of the circle with radius $R^{\min}$. If either of the FQED returns true, we have $B=$ True, and there exists at least one eigenvalue in the region covered by the outer (green) disks. So $R^{\max}$ is updated. Otherwise, we have $B=$ False. In this case,  all of the eigenvalues are outside the region covered by the inner disks, so $R^{\min}$ is updated. 

We then estimate the runtime for $\mathscr{S}_{\text{ring}}$. Because $\nu(r)=O(r^{m_{\max}}/K)$, the number of FQEDs is $|\mathcal{N}_{\text{ring}}(R_{\text{a}},\nu(r))|=\tilde{O}(R_{a}Kr^{-m_{\max}})$. Each query to the FQED has runtime 
$
\tilde{\mathcal{O}}(\gamma^{-1}(\nu(r)^{-1}(C_{\text{be}}+n)+C_{\text{sp}}))=\tilde{\mathcal{O}}(\gamma^{-1}(Kr^{-m_{\max}}(C_{\text{be}}+n)+C_{\text{sp}})).$
Multiplying the total number of FQEDs, we obtain the runtime given in Lemma.~\ref{lm:rg_rt}.

\section{Eigenvalue problem with further simplifications}\label{sec:stp2}
\subsection{Eigenvalue problem without constrains}\label{sec:stp2}

\begin{figure}[t]
    \centering
          \includegraphics[width=.8\columnwidth]{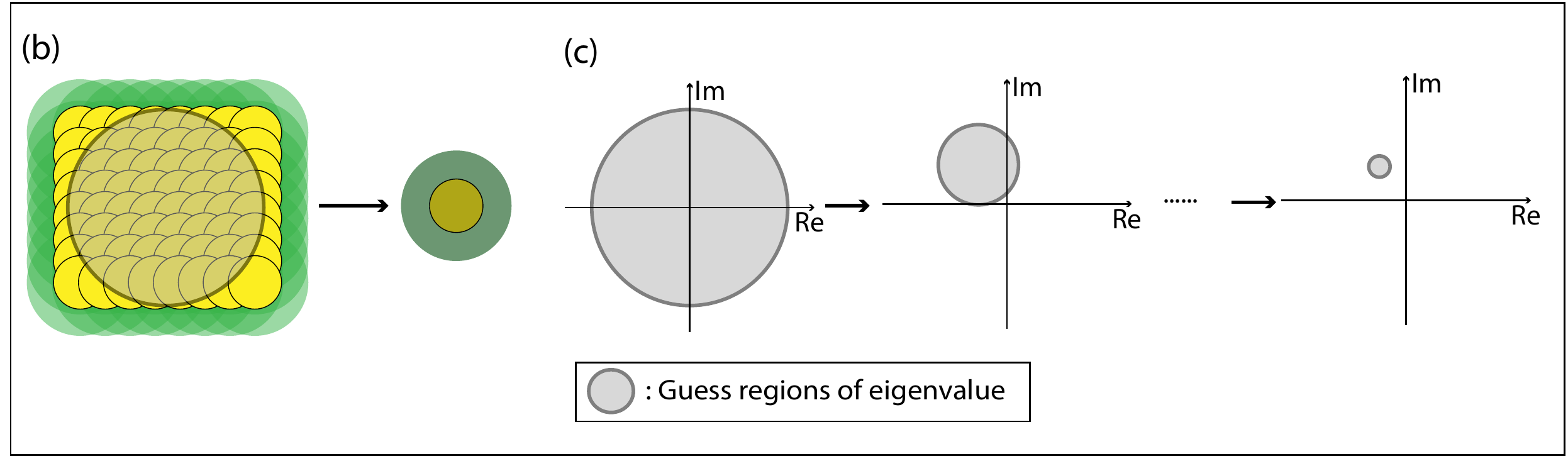}
       \caption{Sketch of Algorithm.~\ref{alg:21} for solving Problem.~\ref{def:1f}. Yellow and green disks represents the inner and outer disks of FQEDs (see Fig.~\ref{fig:alg0}(a) in main text). The initial and updated guess region is enclosed by grey circles. Once an FQED returns \texttt{True}, the guess region is updated. (c) The guess region is updated iteratively until its area is sufficiently small.}
       \label{fig:1}
\end{figure}

In this section, we consider the following problem for general matrices that are not necessarily to be diagonalizable. 
$\\$
\setcounter{problem}{2}
\begin{problem}[]\label{def:1f}
Given a square matrix $A$ with $\|A\|\leqslant1$ and accuracy $\varepsilon\in(0,1)$. The goal is to output an eigenvalue estimation $\lambda'$, such that $\min_{\lambda_j}|\lambda'-\lambda_j|\leqslant\varepsilon$, where $\lambda_j$ are eigenvalues of $A$.
\end{problem}
$\\$
Our solution is summarized as a pseudo-code in Algorithm.~\ref{alg:12}. 
Our strategy of solving Problem.~\ref{def:1f} is to iteratively shrink the region in which there are at least one eigenvalue in it.  Our method contains $J=\lceil\log_2(1/\varepsilon)\rceil$ steps, and the process of each step is illustrated in Fig.~\ref{fig:1}b (see also Algorithm.~\ref{alg:21}). Suppose that before the $j$th step, there are at least one eigenvalue in the region $\mathcal{D}(\lambda_{\text{gss}},D)$, where $\mathcal{D}(a, b) = \{ x \in \mathbb{C} \,|\, |x - a| < b \}$. At this step, we shrink the radius of such confident region from $D$ to $D/2$. This is achieved by introducing a set of FQEDs, whose inner disks cover $\mathcal{D}(\lambda_{\text{gss}},D)$. This ensures that at least one of the FQEDs returns ``True''. Another restriction is that the outer disk of each FQED has a radius $D/2$. In this way, once we obtain an output ``True'', at least one eigenvalue is in the region $\mathcal{D}(\lambda_{\text{gss}}',D/2)$, where $\lambda_{\text{gss}}'$ is the center of such FQED with output ``True''. 

Note that in Algorithm.~\ref{alg:21}, we have introduced a set of points $\mathcal{N}_{\text{net}}(\lambda_{\text{gss}},D,m_{\max})$. It represents the centers of all FQEDs satisfying the criteria above. Equivalently, we have
\begin{align}
\mathcal{D}(\lambda_{\text{gss}},D)&\subset\bigcup_{\mu\in\mathcal{N}_{\text{net}}(\lambda_{\text{gss}},D,m_{\max})}\mathcal{D}(\mu,\nu(D/2)).
\end{align}
The threshold of each FQED is $O(D^{m_{\max}}/K)$. So the complexity of each query to FQED is  $$\tilde{\mathcal{O}}(\gamma^{-1} (KD^{-m_{\max}}(C_{\text{be}}+n)+C_{\text{sp}})).$$ 
The inner disk of each FQED has area $O(D^{2m_{\max}}/K^2)$, so totally $O(K^2D^{-2m_{\max}+2})$ number of FQEDs is required to cover  $\mathcal{D}(\lambda_{\text{gss}},D)$. Therefore, the complexity for each step is 
$$\tilde{\mathcal{O}}(\gamma^{-1} (K^3D^{-3m_{\max}+2}(C_{\text{be}}+n)+K^2D^{-2m_{\max}+2}C_{\text{sp}})).$$
In Algorithm.~\ref{alg:12}, the algorithm contains totally $J=\lceil\log_2(1/\varepsilon)\rceil$ steps, and we have $D\geqslant\varepsilon$. So the total complexity is 
 $$\tilde{\mathcal{O}}(\gamma^{-1} (K^3\varepsilon^{-3m_{\max}+2}(C_{\text{be}}+n)+K^2D^{-2m_{\max}+2}C_{\text{sp}})).$$
 For diagonalizable matrix with $m_{\max}=1$, the runtime reduces to 
  $$\tilde{\mathcal{O}}(K^2\gamma^{-1} (K\varepsilon^{-1}(C_{\text{be}}+n)+C_{\text{sp}})).$$
  Comparing to line gap problem in the main text, the runtime with respect to accuracy reduces from $\mathcal{O}(\varepsilon^{-2})$ to $\mathcal{O}(\varepsilon^{-1})$, thus achieving the Heisenberg limit. 

\begin{algorithm} [H]
\caption{Quantum eigenvalue searching for Problem~\ref{def:1f}. }  
\label{alg:12}  
\begin{algorithmic}%[1]
\STATE $D\leftarrow1$%, $\delta'\leftarrow\delta/\lceil\log_2(D/\varepsilon)\rceil$
\STATE \textbf{while} $D>\varepsilon$:
\STATE \quad $\lambda_{\text{gss}}\leftarrow\mathscr{R}\left(\lambda_{\text{gss}},D\right)$
\STATE \quad $D\leftarrow D/2$:
\STATE \textbf{end while} 
\STATE \textbf{return} $\lambda_{\text{gss}}$ 
\end{algorithmic} 
\end{algorithm}

\begin{algorithm} [H]
\caption{$\mathscr{R}(\lambda_{\text{gss}},D)$ }  
\label{alg:21}  
\begin{algorithmic}%[1]
%\STATE $\delta'\leftarrow \delta/|\mathcal{N}_{\text{net}}(\lambda_{\text{gss}},D,m_{\max})|$
\STATE \textbf{for all} $\mu\in\mathcal{N}_{\text{net}}(\lambda_{\text{gss}},D,m_{\max})$:

%\STATE \quad\textbf{if} $m_{\max}=1$:
%\STATE \quad\quad $B\leftarrow O_C\left(\mu,D/4K\right)$
%\STATE \quad\textbf{else if}: $m_{\max}>1$:
%\STATE \quad\quad $B\leftarrow O_C(\mu,2\nu(D/2))$
%\STATE \quad\textbf{end if}
\STATE \quad\quad $B\leftarrow O_C(\mu,\nu(D/2))$
\STATE \quad \textbf{if} $B = $ True:
\STATE \quad\quad \textbf{break for}
\STATE \quad \textbf{end if}
\STATE \textbf{end for} 
\STATE \textbf{return} $\mu$ 
\end{algorithmic} 
\end{algorithm}

\subsection{Real eigenvalue cases}\label{sec:real}
If we are promised that all eigenvalues are real and matrices are diagonalizable (i.e. $m_{\max}=1$), the search region of eigenvalue becomes the segment $[-1,1]$ in real axis. In this section, we discuss how solutions to eigenproblems can be simplified. 

\subsubsection{Real eigenvalue without constrains}\label{sec:real1}
\begin{figure}[H]
    \centering
          \includegraphics[width=1\columnwidth]{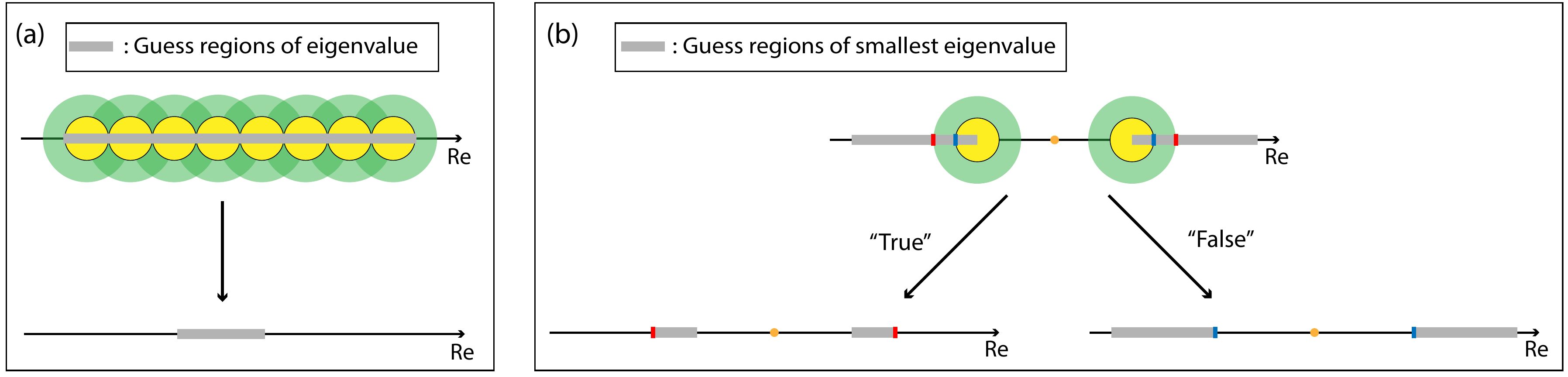}
       \caption{Eigenvalue searching protocols with promised that matrix is diagonalizable and all eigenvalues are real. (a) Sketch of Algorithm.~\ref{alg:real12} for shrinking the range of eigenvalue searching for Problem~\ref{def:1f}. The initial and updated guess region is marked by grey lines. (b) Sketch of Algorithm.~\ref{alg:real_gap_2} for shrinking the range of eigenvalue for Problem~\ref{prob:real}. The initial guess regions are two grey lines. The updated guess regions are two segments with one side marked by red (one of the FQEDs returns \texttt{True}) or blue (all of the FQEDs return \texttt{False}) colors.}
       \label{fig:real}
\end{figure}

Similar to the general case in Sec.~\ref{sec:stp2}, we use a divided-and-conquer strategy and the full algorithm is provided in Algorithm.~\ref{alg:real11}. Before each iteration, the guess region is $[\lambda_{\text{gss}}-D,\lambda_{\text{gss}}+D]$ (initially, we have $\lambda_{\text{gss}}=0$ and $D=1$). After querying $\mathscr{R}_{\text{real}}(\lambda_{\text{gss}},D,\delta)$ defined in Algorithm,~\ref{alg:real12}, $\lambda_{\text{gss}}$ is updated, and $D\rightarrow D/2$. Compared to the $\mathscr{R}$ for general case, the main difference is that $\mathscr{R}_{\text{real}}$ only need to cover segment $[\lambda_{\text{gss}}-D,\lambda_{\text{gss}}+D]$ in real axis with the inner disk of FQED, instead of the entire disk $\mathcal{D}(\lambda_{\text{gss}},D)$. See also Fig.~\ref{fig:real} (a) for illustration.  

In Algorithm.~\ref{alg:real11}, $\mathscr{R}_{\text{real}}$ requires $\mathcal{O}(K)$ queries to FQED, and each query has complexity $\tilde{\mathcal{O}}(\gamma^{-1}(KD^{-1}(C_{\text{be}}+n)+C_{\text{sp}}))$.  
Because $D\geqslant\varepsilon$, and Algorithm.~\ref{alg:real11} has totally $\mathcal{O}(\text{log}(\varepsilon^{-1}))$ queries to $\mathscr{R}_{\text{real}}$,  
we have the following result. \newline
\begin{theorem}
Promised that $\lambda_j\in \mathbb{R}$ for all eigenvalues $\lambda_j$ and $m_{\max}=1$, with arbitrarily high success probability, Problem~\ref{def:1f} can be solved with quantum gate count
\begin{align}\label{eq:search}
\tilde{\mathcal{O}}(K\gamma^{-1}(K\varepsilon^{-1}(C_{{\rm be}}+n)+C_{{\rm sp}}))
\end{align}
\end{theorem}

\subsubsection{Real eigenvalue case for line/point gap problem}\label{sec:real2}
When eigenvalues are promised to be real and diagonalizable, both point gap and line gap problems 
reduce to the following. \newline

\begin{problem}\label{prob:real}
Given a diagonalizable matrix $\|A\|\leqslant1$, promised that $\lambda_j\in\mathbb{R}$  for all eigenvalues $\lambda_j$.
Let $g\equiv\min_{\lambda_j\neq P}\big|\lambda_j-P\big|$ and
$\mathcal{S}\equiv\left\{\lambda_j: |\lambda_j|\in[g,g+\varepsilon]\right\}$ for some accuracy $\varepsilon\in(0,1)$. The goal is to output an eigenvalue estimation $\lambda'$, such that  $|\lambda'-\lambda_j|\leqslant\varepsilon$ for some $\lambda_j\in\mathcal{S}$.\newline
\end{problem}

Note that similar to Section.~\ref{sec:spg}, we have set the reference point as $0$. Our solution to Problem.~\ref{prob:real} is summarized in Algorithm.~\ref{alg:real_gap_1}. 
The main idea is similar to the general case. At each iteration, we are initially promised that $g'\in[R^{\min}_{j},R^{\max}_{j}]$ and we shrink this range by querying a subroutine $\mathscr{S}_{\text{real}}$ defined in Algorithm.~\ref{alg:real_gap_2}. Different from $\mathscr{S}$, the subroutine $\mathscr{S}_{\text{real}}$ uses only two queries to the FQEDs because we only need to cover the corresponding segment in real axis. See Fig.~\ref{fig:real}(b) for illustration. Algorithm.~\ref{alg:real_gap_1} contains two substages as will be discussed below.

Substage 1 corresponds to lines 2-4 in Algorithm.~\ref{alg:real_gap_1}, we begin with analysing the maximal number of iterations required, denoted as $J$. We consider the worst case when we always have $B=$ False. In this case, we have $R^{\min}_{j+1}=R^{\min}_j(1+1/(2K))$, and therefore $R^{\min}_{j}=\varepsilon(1+1/(2K))^j$. Substage 1 terminates whenever $R^{\min}_{j}\leqslant1/2$, so we have $\varepsilon(1+1/(2K))^J=\Theta(1)$. Accordingly, we have
\begin{align}
J=\tilde{\mathcal{O}}\left(\frac{\log((2\varepsilon)^{-1})}{\log(1+1/(2K))}\right)=\tilde{\mathcal{O}}(K\log(\varepsilon^{-1})).
\end{align}
The threshold of each FQED is  bounded by $\Omega(\varepsilon)$, so the total complexity of substage 1 is  
\begin{align}
J\times\tilde{O}(\gamma^{-1}(K\varepsilon^{-1}(C_{\text{be}}+n)+C_{\text{sp}})=\tilde{O}(\gamma^{-1}(K^2\varepsilon^{-1}(C_{\text{be}}+n)+KC_{\text{sp}}).
\end{align}

Substage 2 corresponds to lines 5-7 in Algorithm.~\ref{alg:real_gap_1}. We first define $\Delta_j=R^{\max}_j-R^{\min}_j$. Again, we first analyse the maximal number of iterations in this substage, denoted as $J'$. In the worst case, we always have $B=$ False, which gives $\Delta_{j+1}=\Delta_j(1-1/(4K))$. Suppose substage 2 begins with $j=j'$ and we denote $\Delta=\Delta_{j'}$. Then, we have $\Delta_{j}=\Delta(1-1/(4K))^{j-j'}$. Substage 2 terminates whenever $\Delta_j\leqslant\varepsilon$, so we have $\Delta(1-1/(4K))^{J'}=\Theta(\varepsilon)$. Accordingly, we have 
\begin{align}
J'={\mathcal{O}}\left(\frac{\log(\varepsilon/\Delta)}{\log(1-1/(4K))}\right)={\mathcal{O}}(-K\log(\varepsilon/\Delta))=\tilde{\mathcal{O}}(K).
\end{align}
The runtime of each FQED is still $\Omega(\varepsilon)$. Combining the complexity of both stages, we have the following theorem. 

\begin{theorem}\label{th:7}
With arbitrarily high success probability, Problem~\ref{prob:real} can be solved with quantum gate count
\begin{align}
\tilde{O}(\gamma^{-1}(K^2\varepsilon^{-1}(C_{{\rm be}}+n)+KC_{{\rm sp}})).\notag
\end{align}
\end{theorem}

\subsection{Dissipation of open quantum system: Liouvillian gap}
%\red{https://arxiv.org/pdf/2202.12591.pdf}

The dynamic of a close quantum system is described by Schrodinger's equation with an Hermitian Hamiltonian.  When the system to be studied has interactions with its environment, however, the evolution goes  beyond Hermiticity. This type of open quantum system can be modeled by the Lindblad master equation~\cite{Breuer.02}. Under Markov approximation, the evolution of a quantum state described by density matrix $\rho$ can generally be expressed as
\begin{align}\label{eq:slin1}
\dot{\rho}=\mathcal{L} (\rho)\equiv   -i[H, \rho]+\sum_\mu\left(-\frac{1}{2}L_\mu^{\dagger} L_\mu \rho-\frac{1}{2}\rho L_\mu^{\dagger} L_\mu+L_\mu \rho L_\mu^{\dagger}\right) 
\end{align}
for some Hermitian Hamiltonian $H$ and dissipators $L_\mu$ which are not necessarily to be Hermitian. 
\subsubsection{Vectorization and Block encoding of Liouvillian}

To facilitate the discussion, we consider the case when both $H$ and dissipators $L_\mu$ can be decomposed in the form of
\begin{align}
H&=\sum_{j}\alpha_{j}V_{0,j},\label{eq:hlcu}\\
L_{\mu}&=\sum_j\sqrt{\alpha_{\mu,j}}V_{\mu,j},\label{eq:llcu}
\end{align}
where $\alpha_j,\beta_{j,\mu}>0$. ``$\sqrt{\;\;}$'' in Eq.~\eqref{eq:llcu} is to ensure that the Hermitian and non-Hermitian terms in Eq.~\eqref{eq:slin1} have the same units. $V_j$, $V_{k,\mu}$ are unitaries that can be implemented efficiently on quantum devices. Here, as an example, we only focus on qubit system. By abuse of notations, we let $I,X,Y,Z$ be the single-qubit identity and Pauli operators in this section. We assume that $V_{\mu,j}\in\mathbb{P}^{\otimes n}$ is $n$-qubit Pauli string up to a phase, i.e. $\mathbb{P}=\{\pm I,\pm X,\pm Y,\pm Z, \pm iI,\pm iX,\pm iY,\pm iZ\}$. We also define a normalization factor  $C=\sum_{\mu=0}\sum_{j}\alpha_{\mu,j}$ and assume that $C=\text{poly}(n)$. This model covers most of the dissipative many-body systems of interest. Several examples of the dissipation terms are provided in Table.~\ref{tab:s1}.

\begin{table}[h]
\caption{Correspondence between dissipators and the LCU of the vectorized form. \label{tab:s1}}
\begin{ruledtabular}
\begin{tabular}{lll}
Dissipation type &Dissipators& $\tilde{\mathcal{L}}_{\mu}$\\
\hline
Dephasing&$\eta (Z\rho Z-\rho)$&$\eta(Z\otimes Z-I\otimes I)$\\
Depolarization&$\frac{\eta}{3} (X\rho X+Y\rho Y+Z\rho Z-3\rho)$&$\frac{\eta}{3}(X\otimes X+Y\otimes Y+Z\otimes Z-3I\otimes I)$\\
Damping&$\eta (\sigma^{-}\rho \sigma^{+}-\sigma^{+}\sigma^{-}\rho-\rho\sigma^{+}\sigma^{-})$&$\frac{\eta}{4}(X\otimes(X+iY)+Y\otimes(iX-Y)-Z\otimes I- I\otimes(Z+2I))$\\
\end{tabular}
\end{ruledtabular}
\end{table}

By performing vectorization, Eq.~\eqref{eq:slin1} is equivalent to  $\dot{\tilde\rho}=\tilde{\mathcal{L}} \cdot\tilde \rho$ with 
\begin{align}
\tilde{\rho}=\sum_{j, k} \rho_{j k}|j\rangle \otimes|k\rangle,
\end{align}
where $\rho_{j k}$ is the element of $\rho$ at the $j$th row and $k$th column, and
\begin{subequations}\label{eqs:Lvec}
\begin{align}
\tilde{\mathcal{L}}&=\sum_{\mu}\tilde{\mathcal{L}}_\mu,\\
\tilde{\mathcal{L}}_0&=-i H \otimes I^{\otimes n}+i I^{\otimes n} \otimes H^T,\\
\tilde{\mathcal{L}}_{\mu\geqslant1}&=\left(L_\mu \otimes L_\mu^*-\frac{1}{2}L_\mu^{\dagger} L_\mu \otimes I^{\otimes n}-\frac{1}{2}I^{\otimes n} \otimes L_\mu^T L_\mu^*\right).
\end{align}
\end{subequations}
Accordingly, we have
\begin{subequations}
\begin{align}
\tilde{\mathcal{L}}_0&=\sum_{j}\alpha_{0}(V_{j}\otimes I^{\otimes n})+i\alpha_{j}(I^{\otimes n}\otimes V_{j}^{T}),\\
\tilde{\mathcal{L}}_{\mu\geqslant1}&=\sum_{j,k}\sqrt{\alpha_{\mu,j}\alpha_{\mu,k}}\left(V_{\mu,j}\otimes V_{\mu,k}\right)-\frac{1}{2}\sqrt{\alpha_{\mu,j}\alpha_{\mu,k}}\left(V_{\mu,j}^\dag V_{\mu,k}\otimes I^{\otimes n}\right)-\frac{1}{2}\sqrt{\alpha_{\mu,j}\alpha_{\mu,k}}\left(I^{\otimes n}\otimes V_{\mu,j}^{T}V_{\mu,k}^{*}\right).
\end{align}
\end{subequations}
Due to the following relations,
\begin{subequations}
\begin{align}
&I^*=I, X^*=X, Y^*=-Y, Z^*=Z,\\
&I^T=I, X^T=X, Y^T=-Y, Z^T=Z,
\end{align}
\end{subequations}
$\tilde{\mathcal{L}}$ is also a linear combination of Pauli strings with normalization factor 
\begin{align}
\tilde C&=2\left(\sum_{j}\alpha_{j}+\sum_{\mu}\sum_{j,k}\sqrt{\alpha_{\mu,j}\alpha_{\mu,k}}\right)\notag\\
&\leqslant2\left(\sum_{j}\alpha_{j}+\sum_{\mu}\sum_{j}\alpha_{\mu,j}\right)\notag\\
&=2C.
\end{align}
In other words, we can express the rescaled, vectorized Liouvillian in the form of 
\begin{align}
\tilde{\mathcal{L}}/\tilde C=\sum_{j=0}^{J}\beta_{j}u(j)\label{eq:sbl}
\end{align}
for some $\sum_{j}\beta_j=1$, $J=\text{poly}(n)$, $u_j\in\mathbb{P}^{\otimes n}$ and $\tilde C=O(\text{poly}(n))$. 

We then show how to perform block encoding of Eq.~\eqref{eq:sbl} based on the linear combination of unitaries~\cite{Low.19}. Let $G|0\rangle=\sum_j\sqrt{\beta_j}|j\rangle$ and $\text{Select}(u)=\sum_{j}|j\rangle\langle j|\otimes u(j)$, it can be verified that
\begin{align}
&\left(\langle0|\otimes I^{\otimes n}\right)\left(G^\dag\otimes I^{\otimes n}\right)\text{Select}(u)\left(G\otimes I^{\otimes n}\right)\left(|0\rangle\otimes I^{\otimes n}\right)=\tilde{\mathcal{L}}/\tilde{C}.
\end{align}

\noindent So the quantum circuit below is the block encoding of $\tilde{\mathcal{L}}/\tilde{C}$, which can be constructed with gate count 
$C_{\text{be}}=\mathcal{O}(Jn)$~\cite{Zhang.24}.
\begin{center}
\begin{quantikz}
&\qw\qwbundle{}& \gate{G}& \gate[2]{\text{Select}(u)} & \gate{G^\dag}&\qw \\
&\qw\qwbundle{}&    \qw    &\qw      & \qw&\qw
\end{quantikz}
\end{center}

\subsubsection{Liouvillian gap}
As mentioned in the main text, Liouvillian gap (LG) is defined as the distance between imaginary axis and the eigenvalues closest to it excluding those in the imaginary axis. So the LG of $\tilde{\mathcal{L}}/\tilde C$ is equivalent to the line gap problem with reference line $\bm{L}$ the imaginary axis. To achieve accuracy $\varepsilon$ of $\tilde{\mathcal{L}}$, we should achieve accuracy $\varepsilon/\tilde C$ of $\tilde{\mathcal{L}}/\tilde C$. Because $\tilde C$ is upper bounded by $2C$, we have the following result. \newline

\begin{theorem}\label{th:lvg}
Given a Lindblad master equation described by Eq.~\eqref{eq:slin1}-~\eqref{eq:llcu}, and promise that $\mathcal{L}$ is diagonalizable. With arbitrarily high success probability, the LG can be estimated to accuracy $\varepsilon$ with gate count $$\tilde {\mathcal{O}}(KC\gamma^{-1}(K^2C^2\varepsilon^{-2}Jn+C_{{\rm sp}})).$$ 
\end{theorem}

\subsection{non-Hermitian Hamiltonian: symmetry breaking witness}

\subsubsection{Shrodinger equation with non-Hermitian Hamiltonian}
Effective non-Hermitian Hamiltonian has been widely used in studying open quantum physics, which may emerge from different backgrounds. For completeness of our discussion, we introduce one of the most typical derivations from the short-time limit of Lindblad master equation~\cite{Dalibard.92, Carmichael.93,Nakagawa.18,Song.19}.

The dissipation terms of the Lindblad master equation, i.e. Eq.~\eqref{eq:slin1}, can be separated into two parts. The first part $\mathcal{L}_{\text{con}}(\rho)=\sum_{\mu}-\frac{1}{2}L_\mu^{\dagger} L_\mu \rho-\frac{1}{2}\rho L_\mu^{\dagger} L_\mu$ is called \textit{continuous} dissipation terms, and the second part $\mathcal{L}_{\text{jump}}(\rho)=\sum_\mu L_\mu \rho L_\mu^{\dagger}$ is called \textit{quantum jump} terms. When the evolution time $\tau$ is relatively small, for example, $\tau<\|L_{\mu}L_{\mu}^\dag\|$, the jump term can be neglected and the evolution reduces to
\begin{align}\label{eq:slin2}
\dot{\rho}= -i[H, \rho]+\sum_\mu\left(-\frac{1}{2}L_\mu^{\dagger} L_\mu \rho-\frac{1}{2}\rho L_\mu^{\dagger} L_\mu\right). 
\end{align}
We define 
\begin{align}
H_{\text{eff}}=H-\frac{1}{2}i\sum_\mu L_\mu^{\dagger} L_\mu,
\end{align} 
and Eq.~\eqref{eq:slin2} is equivalent to

\begin{align}\label{eq:slin3}
\dot{\rho}&=-i\left(H_{\text{eff}}\rho -\rho H_{\text{eff}}^\dag\right).
\end{align} 
Instead of density matrix, we may characterize the quantum state with an \textit{unnormalized} wavefunction $|\psi\rangle$. It can be verified that when $\rho=|\psi\rangle\langle\psi|$, Eq.~\eqref{eq:slin2} is equivalent to the following non-Hermitian Schrodinger equation
\begin{align}\label{eq:slin3}
\dot{|\psi\rangle}=-iH_{\text{eff}}|\psi\rangle.
\end{align} 
So the system can be characterized by the effective Hamiltonian $H_{\text{eff}}$, which is in general non-Hermitian.

\subsubsection{Spectrum reality and spontaneous symmetry breaking }

Following the conventions in~\cite{Bender.98,Bender.99,Khare.00}, we let $\mathcal{T}$ be the antilinear complex-conjugation operator which acts as the time-reversal operator, i.e. $\mathcal{T}|\psi\rangle=|\psi\rangle^*$. Note that because $\mathcal{T}^2=I$, we have $\mathcal{T}^{-1}=\mathcal{T}$. Here, we consider a slightly more general case than the parity-time symmetry. More specifically, we let $S$ be an arbitrary invertible matrix, and say that a matrix $H$ has $S\mathcal{T}$-symmetry if 

\begin{align}
 H=S\mathcal{T} H(S\mathcal{T})^{-1}.
\end{align}
Suppose $H|v\rangle=\lambda|v\rangle$, we have
\begin{align}
HS\mathcal{T}|v\rangle=S\mathcal{T} H\mathcal{T}^{-1}S^{-1}(S\mathcal{T}|v\rangle)=S\mathcal{T} H|v\rangle=S (\lambda^* |v\rangle^*)= \lambda^* S|v\rangle^*=\lambda^* S\mathcal{T}|v\rangle.
\end{align}
So $S\mathcal{T}|v\rangle$ is also an eigenvector of $H$ with eigenvalue $\lambda^{*}$. If $\lambda$ is real, $S\mathcal{T}|v\rangle$ and $|v\rangle$ are linearly dependent, and hence $|v\rangle$ is invariant under the transformation of $S\mathcal{T}$, i.e. preserves the $S\mathcal{T}$-symmetry. On the other hand, whenever $\lambda$ is a complex value, $S\mathcal{T}|v\rangle$ and $|v\rangle$ have different eigenvalues and hence linearly independent. So for $|v\rangle$, the $S\mathcal{T}$-symmetry is spontaneously broken. Therefore, the complex eigenvalue serves as a witness for $S\mathcal{T}$-symmetry breaking~\cite{Mostafazadeh.02,Mostafazadeh.02_,Li.22,Melkani.23}. In below, we discuss the search for complex eigenvalues with quantum computing.

\begin{figure}[h]
    \centering
          \includegraphics[width=1\columnwidth]{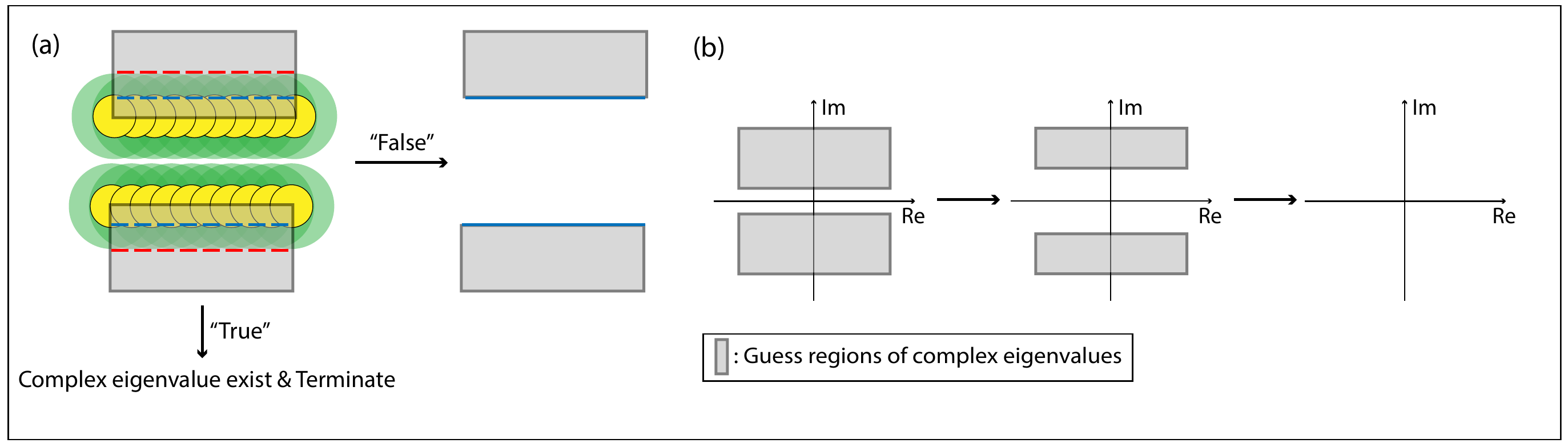}
       \caption{Solution to Problem.~\ref{prob:pt} for witnessing complex eigenvalues. (a) Sketch of each iteration in searching complex eigenvalues. If one of the FQEDs returns \texttt{True}, we terminate the algorithm and conclude the witness of complex eigenvalues. If all FQEDs return \texttt{False}, we shrink the search region. (b) If no complex eigenvalues are witnessed, the search region is updated iteratively until it vanishes.}
       \label{fig:pt}
\end{figure}

\subsubsection{Quantum computing witness of spontaneous symmetry breaking}

While the complex eigenvalue serves as a witness of symmetry breaking, in practice, we allow an accuracy $\varepsilon$. More specifically, we consider the following problem. \newline

\begin{problem}\label{prob:pt}
Given a square and diagonalizable matrix $A$ with $\|A\|\leqslant1$, and $\varepsilon\in(0,1)$. Output ``True'' if there exist eigenvalue $\lambda_j$ satisfying $|\text{Im}(\lambda_j)|\geqslant\varepsilon$; output ``False'' if all eigenvalues satisfy $|\text{Im}(\lambda_j)|\leqslant\varepsilon/2$. For other scenarios, output either ``True'' or ``False''. \newline
\end{problem}

In the ``True'' case, the corresponding matrix is in the symmetry broken phase. In the ``False'' case, however,  both broken and unbroken phase are possible. Therefore, Problem.~\ref{prob:pt} serves as  a \textit{witness} of spontaneous symmetry breaking. 
  
Note that we have also assumed that $A$ is diagonalizable, although the generalization to defective case is possible. Our solution to Problem.~\ref{prob:pt} is given in Algorithm.~\ref{alg:ptg1} in Sec.~\ref{sec:pcpt}. As sketched in Fig.~\ref{fig:pt}, at each iteration, we are confident that there are no eigenvalues with imaginary part $\text{Im}(\lambda_j)\in[\varepsilon,b]$. We update $b$ iteratively by querying a set of FQEDs. The algorithm is terminated whenever a FQED has output ``True''. In this case, we witness a complex eigenvalue. On the other hand, if $b$ increases consistently until $b\geqslant1$ (notice that we always have $|\lambda_j|\leqslant1$), we judge that no complex eigenvalues are witnessed, and terminate the algorithm. 
The runtime of Algorithm.~\ref{alg:ptg1} is similar to the one for point gap and line gap estimations. So we have the following theorem. \newline

\begin{theorem}
With arbitrarily high success probability, Problem.~\ref{prob:pt} can be solved with runtime
\begin{align}
\tilde{\mathcal{O}}\left(K^2\gamma^{-1}\varepsilon^{-1} (K\varepsilon^{-1}(C_{\rm be}+n)+C_{\rm sp})\right).\notag
\end{align}
\end{theorem}

\noindent Note that Problem.~\ref{prob:bqp2} in the main text is another version of Problem.~\ref{prob:pt} with a weaker state preparation assumption, which can still be solved efficiently.

\subsection{Markov process: absolute gap and relaxation time}\label{sec:mp}

\begin{figure}[t]
    \centering
          \includegraphics[width=1\columnwidth]{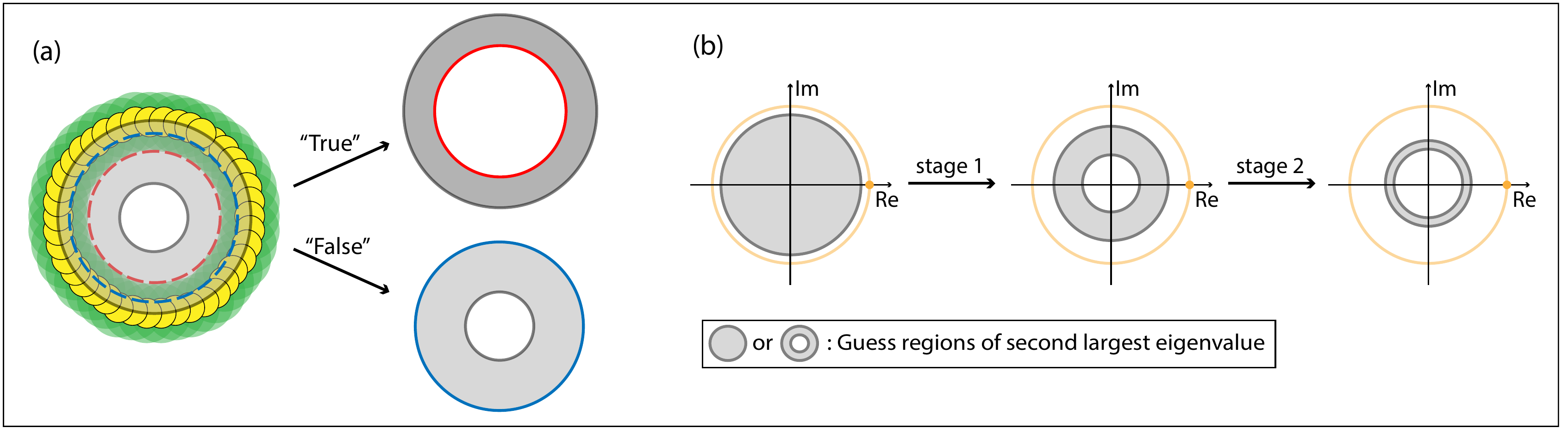}
       \caption{Solution to Problem.~\ref{prob:ag} for estimating absolute gap of Markov process. (a) Sketch of each iteration in estimating the absolute gap of eigenvalues. (b) The guess region of $\lambda_{\max}'$ (second largest absolute value of eigenvalues) is updated iteratively. The reference circle, $\{x:|x|=\lambda_{\max}\}$, is mark with orange color. }
       \label{fig:ag}
\end{figure} 

Markov process is described by the corresponding stochastic matrix. Similar to the previous section, we may  also restrict our discussion to diagonalizable case. The absolute gap problem can be formalized as follows. \newline
\begin{problem}\label{prob:ag}
Let $A$ be a stochastic matrix describing a Markov process. We further assume that $A$ is diagonalizable. Let $g_{\text{ab}}=\lambda_{\max}-\max_{|\lambda_j|\neq|\lambda_{\max}|}|\lambda_j|$ be the absolute gap with promise that $g_{\text{ab}}\geqslant \Delta$ for some $\Delta>\varepsilon$. Output an estimation $g'_{\text{ab}}$ of the absolute gap with accuracy $|g'_{\text{ab}}-g_{\text{ab}}|\leqslant\varepsilon$. \newline
\end{problem}

We assume that a block encoding of matrix $A/\alpha$ is given, with gate complexity $C_{\text{be}}$. Let $\tilde A=A/\alpha$, we have $\|\tilde A\|\leqslant1$, and the maximal eigenvalue of $\tilde A$ is $1/\alpha$.

We let $\lambda'_{\max}$ be the second largest absolute value of the eigenvalues of $\tilde A$. Our algorithm for solving Problem.~\ref{prob:ag} is provided in Algorithm.~\ref{alg:ag0}. We initially have $\lambda'_{\max}\in[0,(1-\Delta)/\alpha]$. In our algorithm, $\lambda'_{\max}$ is updated iteratively.  We suppose that the guess region of $\lambda'_{\max}$ before each iteration is $\lambda'_{\max}\in[R^{\min},R^{\max}]$. As illustrated in Fig.~\ref{fig:ag}, we query a set of FQEDs with centers $\mu\in\mathcal{N}_{\text{ring}}(R^{\text{max}},(1-R^{\text{max}})/K)$. The inner disks of these FQEDs cover the circle with radius $R^{\max}$. If all FQEDs has output ``False'', $R^{\max}$ is updated; if either of the FQED has output ``True'', we update the $R^{\min}$.  This process is encapsulated as an eigenvalue range shrinking subroutine, $\mathscr{S}_{\text{ag}}$, defined in Algorithm.~\ref{alg:erssag}. The update process contains two stages. In stage 1, the radius of the outer disk of each FQED is $1-R^{\max}$. In stage 2, the radius is {$(R^{\max}-R^{\min})/2$} instead.
  
Because $g_{\text{ab}}=1-\lambda'_{\text{max}}\alpha$, to achieve accuracy $\varepsilon$ for $g_{\text{ab}}$, it suffices to achieve accuracy $\varepsilon/\alpha$ for $\lambda'_{\text{max}}$. The complexity of the Algorithm.~\ref{alg:ag0} is similar to Algorithm.~\ref{alg:rg1} for solving Problem.~\ref{prob:pgf}, with $\varepsilon$ replaced by $\varepsilon/\alpha$. 
\newline

\begin{theorem}\label{th:mar}
Problem.~\ref{prob:ag} can be solved with gate count 
\begin{align}
\tilde{\mathcal{O}}\left(K^2\alpha^2\gamma^{-1}\varepsilon^{-1} (K\alpha\varepsilon^{-1}(C_{\rm be}+n)+C_{\rm sp})\right).\notag
\end{align}

\end{theorem}
 
Moreover, we can estimate the relaxation time based on Theorem.~\ref{th:mar}. The relaxation time of the Markov process is defined as $\tau_{\text{rel}}\equiv1/g_{\text{ab}}$. We can define $\tau_{\text{bnd}}=1/\Delta$ as the promised upper bound of $\tau_{\text{rel}}$. Let $\varepsilon_{\text{rel}}$ be the absolute accuracy of $\tau_{\text{rel}}$, we have $\varepsilon_{\text{rel}}\sim\tau^2_{\text{bnd}}\varepsilon$. Accordingly, we can estimate the relaxation time with gate count $\tilde{\mathcal{O}}\left(K^2\alpha^2\tau_{\text{bnd}}^2\gamma^{-1}\varepsilon^{-1} (K\alpha\tau_{\text{bnd}}^2\varepsilon^{-1}(C_{\rm be}+n)+C_{\rm sp})\right)$. While this represents the first efficient complexity result for relaxation time estimation, we believe there is still much room for improvement.

\section{More pseudo codes}

\subsection{pseudo code for solving Problem.~\ref{def:1f} in real and diagonalizable case}

\begin{algorithm} [H]
\caption{Eigenvalue searching for Problem~\ref{def:1f} in real and diagonalizable case. }  
\label{alg:real11}  
\begin{algorithmic}%[1]
\STATE $D\leftarrow1$
\STATE \textbf{while} $D>\varepsilon$:
\STATE \quad $\lambda_{\text{gss}}\leftarrow\mathscr{R}_{\text{real}}\left(\lambda_{\text{gss}},D\right)$
\STATE \quad $D\leftarrow D/2$:
\STATE \textbf{end while} 
\STATE \textbf{return} $\lambda_{\text{gss}}$ 
\end{algorithmic} 
\end{algorithm}

\begin{algorithm} [H]
\caption{$\mathscr{R}_{\text{real}}(\lambda_{\text{gss}},D)$ }  
\label{alg:real12}  
\begin{algorithmic}%[1]
\STATE \textbf{for all} $\mu\in\left\{nD/2K\big|n=0,\pm1,\pm2,\cdots,\pm\lceil 2K\rceil\right\}$:

\STATE \quad $B\leftarrow \text{FQED}\left(\mu,D/4K\right)$

\STATE \quad \textbf{if} $B = $ True:
\STATE \quad\quad \textbf{break for}
\STATE \quad \textbf{end if}
\STATE \textbf{end for} 
\STATE \textbf{return} $\mu$ 
\end{algorithmic} 
\end{algorithm} 

\subsection{pseudo code for solving Problem.~\ref{prob:real} (eigenvalue gap problem for real and diagonalizable case)}

\begin{algorithm} [H]
\caption{Solutions to Problem.~\ref{prob:real}\label{alg:real_gap_1}  }  
\begin{algorithmic}

\STATE  $R_0^{\min}\leftarrow \varepsilon$; $R_0^{\max}\leftarrow 1$; $j\leftarrow1$

\STATE \textbf{while} $R_{j-1}^{\max}-R_{j-1}^{\min}>R_{j-1}^{\min}$: \quad\quad\quad\quad\quad\quad\quad\quad\quad\quad$\#$ substage 1

\STATE \quad $\left(R_{j}^{\min},R_{j}^{\max},S\right)\leftarrow\mathscr{S}_{\text{real}}\left(R_{j-1}^{\min},R_{j-1}^{\max},R_{j-1}^{\min}\right)$

\STATE \textbf{end while}

\STATE \textbf{while} $R_{j-1}^{\max}-R_{j-1}^{\min}>\varepsilon$ \textbf{and} $S\neq0$:   \quad\quad\quad\quad\quad\quad\quad$\#$ substage 2

\STATE \quad $\left(R^{\min}_{j},R^{\max}_{j},S\right)\leftarrow\mathscr{S}_{\text{real}}\left(R_{j-1}^{\min},R_{j-1}^{\max},(R_{j-1}^{\max}-R_{j-1}^{\min})/(2K)\right)$
\STATE \textbf{end while}

\STATE \textbf{return $S\times (R_{j}^{\max}+R_{j}^{\min})/2$}

\end{algorithmic} 
\end{algorithm} 

\begin{algorithm}[H]
\caption{$\mathscr{S}_{\text{real}}\left(R_a,R_b,r\right)$\label{alg:real_gap_2} }  
\begin{algorithmic}

\STATE \quad $B_+\leftarrow \text{FQED}(R_a,r/(2K))$
\STATE \quad $B_-\leftarrow \text{FQED}(-R_a,r/(2K))$
\STATE \quad $B=B_+\vee B_-$
\STATE \quad \textbf{if} $B=$ False:

\STATE \quad\quad $S=0$

\STATE \quad \textbf{else if} $B_+=$ True:

\STATE \quad\quad $S=1$

\STATE \quad \textbf{else if} $B_-=$ True:

\STATE \quad\quad $S=-1$

\STATE \quad \textbf{end if}

\STATE \quad \textbf{if} $B=$ True:

\STATE \quad\quad $\tilde{R}_a\leftarrow R_a$
\STATE \quad\quad $\tilde{R}_b\leftarrow R_b+r$

\STATE \quad \textbf{else}:

\STATE \quad\quad $\tilde{R}_a\leftarrow R_a+r/(2K)$
\STATE \quad\quad $\tilde{R}_b\leftarrow R_b$

\STATE \quad \textbf{end if}

\STATE \quad \textbf{return} $\left(R_a,R_b,S\right)$

\end{algorithmic} 
\end{algorithm} 

\subsection{pseudo code for solving  Problem.~\ref{prob:pt} (complex eigenvalue witness)}\label{sec:pcpt}

\begin{algorithm} [H]
\caption{  Algorithm solving Problem~\ref{prob:pt}}  
\label{alg:ptg1}  
\begin{algorithmic}

\STATE $b\leftarrow\varepsilon$

\STATE \textbf{while} $b< 1$:

\STATE $\Delta_a\leftarrow \lfloor b/(4K)\rfloor $

\STATE \quad\textbf{for} $a\in\{-1,-1+\Delta_a,-1+2\Delta_a,\cdots,1\}$: 

\STATE\quad\quad   $B_+\leftarrow O_C(a+ib,2b/\Delta_a)$
\STATE\quad\quad   $B_-\leftarrow O_C(a-ib,2b/\Delta_a)$
\STATE\quad\quad  \textbf{if} $B_-=$ True or $B_+=$ True:
\STATE\quad\quad\quad \textbf{return True} and \textbf{Terminate}
\STATE\quad\quad  \textbf{end if} 
\STATE \quad\textbf{end for}

\STATE\quad $b\leftarrow b(1+1/(4K))$

\STATE \textbf{end while} 
\STATE \textbf{return False}

\end{algorithmic} 
\end{algorithm}

\subsection{pseudo code for solving  Problem.~\ref{prob:ag} (eigenvalue absolute gap estimation)}

\begin{algorithm} [H]
\caption{ Solution to Problem~\ref{prob:ag}}  
\label{alg:ag0}  
\begin{algorithmic}

\STATE $R_0^{\min}\leftarrow0$; $R_0^{\max}\leftarrow(1-\Delta)/C$; $j\leftarrow1$

\STATE \textbf{while} $ (1/C-R_{j-1}^{\max})/2<1/C-R_{j-1}^{\min}$: $\quad\quad\quad\quad\quad\quad\quad\quad\quad\quad\quad\#$ Stage 1

\STATE \quad $(R_{j}^{\max},R_{j}^{\min})\leftarrow\mathscr{S}_{\text{ag}}(R_{j-1}^{\max},R_{j-1}^{\min}, 1-R_{j-1}^{\max})$

\STATE \quad $j\leftarrow j+1$

\STATE \textbf{end while} 

\STATE \textbf{while} $R_{j-1}^{\max}-R_{j-1}^{\min}>\varepsilon/C$: 
$\quad\quad\quad\quad\quad\quad\quad\quad\quad\quad\quad\quad\quad\quad\;\#$ Stage 2

\STATE \quad $(R_{j}^{\max},R_{j}^{\min})\leftarrow\mathscr{S}_{\text{ag}}\left(R_{j-1}^{\max},R_{j-1}^{\min}, \left(R_{j-1}^{\max}-R_{j-1}^{\min}\right)/2\right)$

\STATE \quad $j\leftarrow j+1$

\STATE \textbf{end while}

\STATE \textbf{return} $1-\left(\left(R_{j-1}^{\min}+R_{j-1}^{\max}\right)/2\right)$

\end{algorithmic} 
\end{algorithm}

\begin{algorithm} [H]
\caption{ $\mathscr{S}_{\text{ag}}(R_{\text{a}},R_{\text{b}},r)$ }  
\label{alg:erssag}  
\begin{algorithmic}

\STATE $\delta'\leftarrow\delta/|\mathcal{N}_{\text{ring}}(R_{\text{a}},r/2K)|$ $\quad\quad$ $\#$ $\mathcal{N}_{\text{ring}}$ is defined in Eq.~\eqref{eq:T1}
\STATE \textbf{for all} $t\in \mathcal{N}_{\text{ring}}(R_{\text{a}},r/2K)$:

\STATE \quad $B\leftarrow \text{FQED}(t,r/2K)$  \quad\quad\quad$\#$ FQED works for the rescaled matrix $\tilde A=A/\alpha$

\STATE \quad \textbf{if} $B=$ True:

\STATE \quad\quad \textbf{break for }

\STATE \quad \textbf{end if}

\STATE \textbf{end for} 

\STATE \textbf{if}  $B=$ True:

\STATE \quad  $\tilde R_{\text{a}}\leftarrow R_{\text{a}}$
\STATE \quad  $\tilde R_{\text{b}}\leftarrow R_{\text{a}}-r$

\STATE \textbf{else if}  $B=$ False:

\STATE \quad $\tilde R_{\text{a}}\leftarrow R_{\text{a}}-r/(4K)$
\STATE \quad  $\tilde R_{\text{b}}\leftarrow R_{\text{b}}$

\STATE \textbf{end if}

\STATE \textbf{return} $\left(\tilde R_{\text{a}}, \tilde R_{\text{b}}\right)$ 

\end{algorithmic} 
\end{algorithm}

%\bibliography{refs_eg}

\begin{thebibliography}{70}%
\makeatletter
\providecommand \@ifxundefined [1]{%
 \@ifx{#1\undefined}
}%
\providecommand \@ifnum [1]{%
 \ifnum #1\expandafter \@firstoftwo
 \else \expandafter \@secondoftwo
 \fi
}%
\providecommand \@ifx [1]{%
 \ifx #1\expandafter \@firstoftwo
 \else \expandafter \@secondoftwo
 \fi
}%
\providecommand \natexlab [1]{#1}%
\providecommand \enquote  [1]{``#1''}%
\providecommand \bibnamefont  [1]{#1}%
\providecommand \bibfnamefont [1]{#1}%
\providecommand \citenamefont [1]{#1}%
\providecommand \href@noop [0]{\@secondoftwo}%
\providecommand \href [0]{\begingroup \@sanitize@url \@href}%
\providecommand \@href[1]{\@@startlink{#1}\@@href}%
\providecommand \@@href[1]{\endgroup#1\@@endlink}%
\providecommand \@sanitize@url [0]{\catcode `\\12\catcode `\$12\catcode
  `\&12\catcode `\#12\catcode `\^12\catcode `\_12\catcode `\%12\relax}%
\providecommand \@@startlink[1]{}%
\providecommand \@@endlink[0]{}%
\providecommand \url  [0]{\begingroup\@sanitize@url \@url }%
\providecommand \@url [1]{\endgroup\@href {#1}{\urlprefix }}%
\providecommand \urlprefix  [0]{URL }%
\providecommand \Eprint [0]{\href }%
\providecommand \doibase [0]{http://dx.doi.org/}%
\providecommand \selectlanguage [0]{\@gobble}%
\providecommand \bibinfo  [0]{\@secondoftwo}%
\providecommand \bibfield  [0]{\@secondoftwo}%
\providecommand \translation [1]{[#1]}%
\providecommand \BibitemOpen [0]{}%
\providecommand \bibitemStop [0]{}%
\providecommand \bibitemNoStop [0]{.\EOS\space}%
\providecommand \EOS [0]{\spacefactor3000\relax}%
\providecommand \BibitemShut  [1]{\csname bibitem#1\endcsname}%
\let\auto@bib@innerbib\@empty
%</preamble>
\bibitem [{\citenamefont {Bender}\ and\ \citenamefont
  {Boettcher}(1998)}]{Bender.98}%
  \BibitemOpen
  \bibfield  {author} {\bibinfo {author} {\bibfnamefont {C.~M.}\ \bibnamefont
  {Bender}}\ and\ \bibinfo {author} {\bibfnamefont {S.}~\bibnamefont
  {Boettcher}},\ }\bibfield  {title} {\bibinfo {title} {Real spectra in
  non-hermitian hamiltonians having p t symmetry},\ }\href@noop {} {\bibfield
  {journal} {\bibinfo  {journal} {Physical review letters}\ }\textbf {\bibinfo
  {volume} {80}},\ \bibinfo {pages} {5243} (\bibinfo {year}
  {1998})}\BibitemShut {NoStop}%
\bibitem [{\citenamefont {Bender}\ \emph {et~al.}(1999)\citenamefont {Bender},
  \citenamefont {Boettcher},\ and\ \citenamefont {Meisinger}}]{Bender.99}%
  \BibitemOpen
  \bibfield  {author} {\bibinfo {author} {\bibfnamefont {C.~M.}\ \bibnamefont
  {Bender}}, \bibinfo {author} {\bibfnamefont {S.}~\bibnamefont {Boettcher}}, \
  and\ \bibinfo {author} {\bibfnamefont {P.~N.}\ \bibnamefont {Meisinger}},\
  }\bibfield  {title} {\bibinfo {title} {Pt-symmetric quantum mechanics},\
  }\href@noop {} {\bibfield  {journal} {\bibinfo  {journal} {Journal of
  Mathematical Physics}\ }\textbf {\bibinfo {volume} {40}},\ \bibinfo {pages}
  {2201} (\bibinfo {year} {1999})}\BibitemShut {NoStop}%
\bibitem [{\citenamefont {Khare}\ and\ \citenamefont
  {Mandal}(2000)}]{Khare.00}%
  \BibitemOpen
  \bibfield  {author} {\bibinfo {author} {\bibfnamefont {A.}~\bibnamefont
  {Khare}}\ and\ \bibinfo {author} {\bibfnamefont {B.~P.}\ \bibnamefont
  {Mandal}},\ }\bibfield  {title} {\bibinfo {title} {A pt-invariant potential
  with complex qes eigenvalues},\ }\href@noop {} {\bibfield  {journal}
  {\bibinfo  {journal} {Physics Letters A}\ }\textbf {\bibinfo {volume}
  {272}},\ \bibinfo {pages} {53} (\bibinfo {year} {2000})}\BibitemShut
  {NoStop}%
\bibitem [{\citenamefont {Delabaere}\ and\ \citenamefont
  {Trinh}(2000)}]{Delabaere.00}%
  \BibitemOpen
  \bibfield  {author} {\bibinfo {author} {\bibfnamefont {E.}~\bibnamefont
  {Delabaere}}\ and\ \bibinfo {author} {\bibfnamefont {D.~T.}\ \bibnamefont
  {Trinh}},\ }\bibfield  {title} {\bibinfo {title} {Spectral analysis of the
  complex cubic oscillator},\ }\href@noop {} {\bibfield  {journal} {\bibinfo
  {journal} {Journal of Physics A: Mathematical and General}\ }\textbf
  {\bibinfo {volume} {33}},\ \bibinfo {pages} {8771} (\bibinfo {year}
  {2000})}\BibitemShut {NoStop}%
\bibitem [{\citenamefont {Mostafazadeh}(2002{\natexlab{a}})}]{Mostafazadeh.02}%
  \BibitemOpen
  \bibfield  {author} {\bibinfo {author} {\bibfnamefont {A.}~\bibnamefont
  {Mostafazadeh}},\ }\bibfield  {title} {\bibinfo {title} {Pseudo-hermiticity
  versus pt symmetry: the necessary condition for the reality of the spectrum
  of a non-hermitian hamiltonian},\ }\href@noop {} {\bibfield  {journal}
  {\bibinfo  {journal} {Journal of Mathematical Physics}\ }\textbf {\bibinfo
  {volume} {43}},\ \bibinfo {pages} {205} (\bibinfo {year}
  {2002}{\natexlab{a}})}\BibitemShut {NoStop}%
\bibitem [{\citenamefont
  {Mostafazadeh}(2002{\natexlab{b}})}]{Mostafazadeh.02_}%
  \BibitemOpen
  \bibfield  {author} {\bibinfo {author} {\bibfnamefont {A.}~\bibnamefont
  {Mostafazadeh}},\ }\bibfield  {title} {\bibinfo {title} {Pseudo-hermiticity
  versus pt-symmetry. ii. a complete characterization of non-hermitian
  hamiltonians with a real spectrum},\ }\href@noop {} {\bibfield  {journal}
  {\bibinfo  {journal} {Journal of Mathematical Physics}\ }\textbf {\bibinfo
  {volume} {43}},\ \bibinfo {pages} {2814} (\bibinfo {year}
  {2002}{\natexlab{b}})}\BibitemShut {NoStop}%
\bibitem [{\citenamefont {El-Ganainy}\ \emph
  {et~al.}(2018{\natexlab{a}})\citenamefont {El-Ganainy}, \citenamefont
  {Makris}, \citenamefont {Khajavikhan}, \citenamefont {Musslimani},
  \citenamefont {Rotter},\ and\ \citenamefont {Christodoulides}}]{Ganainy.18}%
  \BibitemOpen
  \bibfield  {author} {\bibinfo {author} {\bibfnamefont {R.}~\bibnamefont
  {El-Ganainy}}, \bibinfo {author} {\bibfnamefont {K.~G.}\ \bibnamefont
  {Makris}}, \bibinfo {author} {\bibfnamefont {M.}~\bibnamefont {Khajavikhan}},
  \bibinfo {author} {\bibfnamefont {Z.~H.}\ \bibnamefont {Musslimani}},
  \bibinfo {author} {\bibfnamefont {S.}~\bibnamefont {Rotter}}, \ and\ \bibinfo
  {author} {\bibfnamefont {D.~N.}\ \bibnamefont {Christodoulides}},\ }\bibfield
   {title} {\bibinfo {title} {Non-hermitian physics and pt symmetry},\
  }\href@noop {} {\bibfield  {journal} {\bibinfo  {journal} {Nature Physics}\
  }\textbf {\bibinfo {volume} {14}},\ \bibinfo {pages} {11} (\bibinfo {year}
  {2018}{\natexlab{a}})}\BibitemShut {NoStop}%
\bibitem [{\citenamefont {Wu}\ \emph {et~al.}(2019)\citenamefont {Wu},
  \citenamefont {Liu}, \citenamefont {Geng}, \citenamefont {Song},
  \citenamefont {Ye}, \citenamefont {Duan}, \citenamefont {Rong},\ and\
  \citenamefont {Du}}]{Wu.19}%
  \BibitemOpen
  \bibfield  {author} {\bibinfo {author} {\bibfnamefont {Y.}~\bibnamefont
  {Wu}}, \bibinfo {author} {\bibfnamefont {W.}~\bibnamefont {Liu}}, \bibinfo
  {author} {\bibfnamefont {J.}~\bibnamefont {Geng}}, \bibinfo {author}
  {\bibfnamefont {X.}~\bibnamefont {Song}}, \bibinfo {author} {\bibfnamefont
  {X.}~\bibnamefont {Ye}}, \bibinfo {author} {\bibfnamefont {C.-K.}\
  \bibnamefont {Duan}}, \bibinfo {author} {\bibfnamefont {X.}~\bibnamefont
  {Rong}}, \ and\ \bibinfo {author} {\bibfnamefont {J.}~\bibnamefont {Du}},\
  }\bibfield  {title} {\bibinfo {title} {Observation of parity-time symmetry
  breaking in a single-spin system},\ }\href@noop {} {\bibfield  {journal}
  {\bibinfo  {journal} {Science}\ }\textbf {\bibinfo {volume} {364}},\ \bibinfo
  {pages} {878} (\bibinfo {year} {2019})}\BibitemShut {NoStop}%
\bibitem [{\citenamefont {Weidemann}\ \emph {et~al.}(2022)\citenamefont
  {Weidemann}, \citenamefont {Kremer}, \citenamefont {Longhi},\ and\
  \citenamefont {Szameit}}]{Weidemann.22}%
  \BibitemOpen
  \bibfield  {author} {\bibinfo {author} {\bibfnamefont {S.}~\bibnamefont
  {Weidemann}}, \bibinfo {author} {\bibfnamefont {M.}~\bibnamefont {Kremer}},
  \bibinfo {author} {\bibfnamefont {S.}~\bibnamefont {Longhi}}, \ and\ \bibinfo
  {author} {\bibfnamefont {A.}~\bibnamefont {Szameit}},\ }\bibfield  {title}
  {\bibinfo {title} {Topological triple phase transition in non-hermitian
  floquet quasicrystals},\ }\href@noop {} {\bibfield  {journal} {\bibinfo
  {journal} {Nature}\ }\textbf {\bibinfo {volume} {601}},\ \bibinfo {pages}
  {354} (\bibinfo {year} {2022})}\BibitemShut {NoStop}%
\bibitem [{\citenamefont {Zhang}\ \emph {et~al.}(2022)\citenamefont {Zhang},
  \citenamefont {Zhang}, \citenamefont {Lu},\ and\ \citenamefont
  {Chen}}]{Xiujuan.22}%
  \BibitemOpen
  \bibfield  {author} {\bibinfo {author} {\bibfnamefont {X.}~\bibnamefont
  {Zhang}}, \bibinfo {author} {\bibfnamefont {T.}~\bibnamefont {Zhang}},
  \bibinfo {author} {\bibfnamefont {M.-H.}\ \bibnamefont {Lu}}, \ and\ \bibinfo
  {author} {\bibfnamefont {Y.-F.}\ \bibnamefont {Chen}},\ }\bibfield  {title}
  {\bibinfo {title} {A review on non-hermitian skin effect},\ }\href@noop {}
  {\bibfield  {journal} {\bibinfo  {journal} {Advances in Physics: X}\ }\textbf
  {\bibinfo {volume} {7}},\ \bibinfo {pages} {2109431} (\bibinfo {year}
  {2022})}\BibitemShut {NoStop}%
\bibitem [{\citenamefont {Okuma}\ and\ \citenamefont {Sato}(2019)}]{Okuma.19}%
  \BibitemOpen
  \bibfield  {author} {\bibinfo {author} {\bibfnamefont {N.}~\bibnamefont
  {Okuma}}\ and\ \bibinfo {author} {\bibfnamefont {M.}~\bibnamefont {Sato}},\
  }\bibfield  {title} {\bibinfo {title} {Topological phase transition driven by
  infinitesimal instability: Majorana fermions in non-hermitian spintronics},\
  }\href@noop {} {\bibfield  {journal} {\bibinfo  {journal} {Physical review
  letters}\ }\textbf {\bibinfo {volume} {123}},\ \bibinfo {pages} {097701}
  (\bibinfo {year} {2019})}\BibitemShut {NoStop}%
\bibitem [{\citenamefont {Longhi}(2019)}]{Longhi.19}%
  \BibitemOpen
  \bibfield  {author} {\bibinfo {author} {\bibfnamefont {S.}~\bibnamefont
  {Longhi}},\ }\bibfield  {title} {\bibinfo {title} {Topological phase
  transition in non-hermitian quasicrystals},\ }\href@noop {} {\bibfield
  {journal} {\bibinfo  {journal} {Physical review letters}\ }\textbf {\bibinfo
  {volume} {122}},\ \bibinfo {pages} {237601} (\bibinfo {year}
  {2019})}\BibitemShut {NoStop}%
\bibitem [{\citenamefont {Cao}\ and\ \citenamefont {Wiersig}(2015)}]{Cao.15}%
  \BibitemOpen
  \bibfield  {author} {\bibinfo {author} {\bibfnamefont {H.}~\bibnamefont
  {Cao}}\ and\ \bibinfo {author} {\bibfnamefont {J.}~\bibnamefont {Wiersig}},\
  }\bibfield  {title} {\bibinfo {title} {Dielectric microcavities: Model
  systems for wave chaos and non-hermitian physics},\ }\href@noop {} {\bibfield
   {journal} {\bibinfo  {journal} {Reviews of Modern Physics}\ }\textbf
  {\bibinfo {volume} {87}},\ \bibinfo {pages} {61} (\bibinfo {year}
  {2015})}\BibitemShut {NoStop}%
\bibitem [{\citenamefont {El-Ganainy}\ \emph
  {et~al.}(2018{\natexlab{b}})\citenamefont {El-Ganainy}, \citenamefont
  {Makris}, \citenamefont {Khajavikhan}, \citenamefont {Musslimani},
  \citenamefont {Rotter},\ and\ \citenamefont {Christodoulides}}]{el.18}%
  \BibitemOpen
  \bibfield  {author} {\bibinfo {author} {\bibfnamefont {R.}~\bibnamefont
  {El-Ganainy}}, \bibinfo {author} {\bibfnamefont {K.~G.}\ \bibnamefont
  {Makris}}, \bibinfo {author} {\bibfnamefont {M.}~\bibnamefont {Khajavikhan}},
  \bibinfo {author} {\bibfnamefont {Z.~H.}\ \bibnamefont {Musslimani}},
  \bibinfo {author} {\bibfnamefont {S.}~\bibnamefont {Rotter}}, \ and\ \bibinfo
  {author} {\bibfnamefont {D.~N.}\ \bibnamefont {Christodoulides}},\ }\bibfield
   {title} {\bibinfo {title} {Non-hermitian physics and pt symmetry},\
  }\href@noop {} {\bibfield  {journal} {\bibinfo  {journal} {Nature Physics}\
  }\textbf {\bibinfo {volume} {14}},\ \bibinfo {pages} {11} (\bibinfo {year}
  {2018}{\natexlab{b}})}\BibitemShut {NoStop}%
\bibitem [{\citenamefont {Ashida}\ \emph {et~al.}(2020)\citenamefont {Ashida},
  \citenamefont {Gong},\ and\ \citenamefont {Ueda}}]{Ashida.20}%
  \BibitemOpen
  \bibfield  {author} {\bibinfo {author} {\bibfnamefont {Y.}~\bibnamefont
  {Ashida}}, \bibinfo {author} {\bibfnamefont {Z.}~\bibnamefont {Gong}}, \ and\
  \bibinfo {author} {\bibfnamefont {M.}~\bibnamefont {Ueda}},\ }\bibfield
  {title} {\bibinfo {title} {Non-hermitian physics},\ }\href@noop {} {\bibfield
   {journal} {\bibinfo  {journal} {Advances in Physics}\ }\textbf {\bibinfo
  {volume} {69}},\ \bibinfo {pages} {249} (\bibinfo {year} {2020})}\BibitemShut
  {NoStop}%
\bibitem [{\citenamefont {Okuma}\ and\ \citenamefont {Sato}(2023)}]{Okuma.23}%
  \BibitemOpen
  \bibfield  {author} {\bibinfo {author} {\bibfnamefont {N.}~\bibnamefont
  {Okuma}}\ and\ \bibinfo {author} {\bibfnamefont {M.}~\bibnamefont {Sato}},\
  }\bibfield  {title} {\bibinfo {title} {Non-hermitian topological phenomena: A
  review},\ }\href@noop {} {\bibfield  {journal} {\bibinfo  {journal} {Annual
  Review of Condensed Matter Physics}\ }\textbf {\bibinfo {volume} {14}},\
  \bibinfo {pages} {83} (\bibinfo {year} {2023})}\BibitemShut {NoStop}%
\bibitem [{\citenamefont {Dalibard}\ \emph {et~al.}(1992)\citenamefont
  {Dalibard}, \citenamefont {Castin},\ and\ \citenamefont
  {M{\o}lmer}}]{Dalibard.92}%
  \BibitemOpen
  \bibfield  {author} {\bibinfo {author} {\bibfnamefont {J.}~\bibnamefont
  {Dalibard}}, \bibinfo {author} {\bibfnamefont {Y.}~\bibnamefont {Castin}}, \
  and\ \bibinfo {author} {\bibfnamefont {K.}~\bibnamefont {M{\o}lmer}},\
  }\bibfield  {title} {\bibinfo {title} {Wave-function approach to dissipative
  processes in quantum optics},\ }\href@noop {} {\bibfield  {journal} {\bibinfo
   {journal} {Physical review letters}\ }\textbf {\bibinfo {volume} {68}},\
  \bibinfo {pages} {580} (\bibinfo {year} {1992})}\BibitemShut {NoStop}%
\bibitem [{\citenamefont {Carmichael}(1993)}]{Carmichael.93}%
  \BibitemOpen
  \bibfield  {author} {\bibinfo {author} {\bibfnamefont {H.~J.}\ \bibnamefont
  {Carmichael}},\ }\bibfield  {title} {\bibinfo {title} {Quantum trajectory
  theory for cascaded open systems},\ }\href@noop {} {\bibfield  {journal}
  {\bibinfo  {journal} {Physical review letters}\ }\textbf {\bibinfo {volume}
  {70}},\ \bibinfo {pages} {2273} (\bibinfo {year} {1993})}\BibitemShut
  {NoStop}%
\bibitem [{\citenamefont {Nakagawa}\ \emph {et~al.}(2018)\citenamefont
  {Nakagawa}, \citenamefont {Kawakami},\ and\ \citenamefont
  {Ueda}}]{Nakagawa.18}%
  \BibitemOpen
  \bibfield  {author} {\bibinfo {author} {\bibfnamefont {M.}~\bibnamefont
  {Nakagawa}}, \bibinfo {author} {\bibfnamefont {N.}~\bibnamefont {Kawakami}},
  \ and\ \bibinfo {author} {\bibfnamefont {M.}~\bibnamefont {Ueda}},\
  }\bibfield  {title} {\bibinfo {title} {Non-hermitian kondo effect in
  ultracold alkaline-earth atoms},\ }\href@noop {} {\bibfield  {journal}
  {\bibinfo  {journal} {Physical review letters}\ }\textbf {\bibinfo {volume}
  {121}},\ \bibinfo {pages} {203001} (\bibinfo {year} {2018})}\BibitemShut
  {NoStop}%
\bibitem [{\citenamefont {Song}\ \emph {et~al.}(2019)\citenamefont {Song},
  \citenamefont {Yao},\ and\ \citenamefont {Wang}}]{Song.19}%
  \BibitemOpen
  \bibfield  {author} {\bibinfo {author} {\bibfnamefont {F.}~\bibnamefont
  {Song}}, \bibinfo {author} {\bibfnamefont {S.}~\bibnamefont {Yao}}, \ and\
  \bibinfo {author} {\bibfnamefont {Z.}~\bibnamefont {Wang}},\ }\bibfield
  {title} {\bibinfo {title} {Non-hermitian skin effect and chiral damping in
  open quantum systems},\ }\href@noop {} {\bibfield  {journal} {\bibinfo
  {journal} {Physical review letters}\ }\textbf {\bibinfo {volume} {123}},\
  \bibinfo {pages} {170401} (\bibinfo {year} {2019})}\BibitemShut {NoStop}%
\bibitem [{\citenamefont {Dressel}\ \emph {et~al.}(2014)\citenamefont
  {Dressel}, \citenamefont {Malik}, \citenamefont {Miatto}, \citenamefont
  {Jordan},\ and\ \citenamefont {Boyd}}]{Dressel.14}%
  \BibitemOpen
  \bibfield  {author} {\bibinfo {author} {\bibfnamefont {J.}~\bibnamefont
  {Dressel}}, \bibinfo {author} {\bibfnamefont {M.}~\bibnamefont {Malik}},
  \bibinfo {author} {\bibfnamefont {F.~M.}\ \bibnamefont {Miatto}}, \bibinfo
  {author} {\bibfnamefont {A.~N.}\ \bibnamefont {Jordan}}, \ and\ \bibinfo
  {author} {\bibfnamefont {R.~W.}\ \bibnamefont {Boyd}},\ }\bibfield  {title}
  {\bibinfo {title} {Colloquium: Understanding quantum weak values: Basics and
  applications},\ }\href@noop {} {\bibfield  {journal} {\bibinfo  {journal}
  {Reviews of Modern Physics}\ }\textbf {\bibinfo {volume} {86}},\ \bibinfo
  {pages} {307} (\bibinfo {year} {2014})}\BibitemShut {NoStop}%
\bibitem [{\citenamefont {Bukov}\ \emph {et~al.}(2015)\citenamefont {Bukov},
  \citenamefont {D'Alessio},\ and\ \citenamefont {Polkovnikov}}]{Bukov.15}%
  \BibitemOpen
  \bibfield  {author} {\bibinfo {author} {\bibfnamefont {M.}~\bibnamefont
  {Bukov}}, \bibinfo {author} {\bibfnamefont {L.}~\bibnamefont {D'Alessio}}, \
  and\ \bibinfo {author} {\bibfnamefont {A.}~\bibnamefont {Polkovnikov}},\
  }\bibfield  {title} {\bibinfo {title} {Universal high-frequency behavior of
  periodically driven systems: from dynamical stabilization to floquet
  engineering},\ }\href@noop {} {\bibfield  {journal} {\bibinfo  {journal}
  {Advances in Physics}\ }\textbf {\bibinfo {volume} {64}},\ \bibinfo {pages}
  {139} (\bibinfo {year} {2015})}\BibitemShut {NoStop}%
\bibitem [{\citenamefont {Oka}\ and\ \citenamefont {Kitamura}(2019)}]{Oka.19}%
  \BibitemOpen
  \bibfield  {author} {\bibinfo {author} {\bibfnamefont {T.}~\bibnamefont
  {Oka}}\ and\ \bibinfo {author} {\bibfnamefont {S.}~\bibnamefont {Kitamura}},\
  }\bibfield  {title} {\bibinfo {title} {Floquet engineering of quantum
  materials},\ }\href@noop {} {\bibfield  {journal} {\bibinfo  {journal}
  {Annual Review of Condensed Matter Physics}\ }\textbf {\bibinfo {volume}
  {10}},\ \bibinfo {pages} {387} (\bibinfo {year} {2019})}\BibitemShut
  {NoStop}%
\bibitem [{\citenamefont {Poulin}\ and\ \citenamefont
  {Wocjan}(2009)}]{Poulin.09}%
  \BibitemOpen
  \bibfield  {author} {\bibinfo {author} {\bibfnamefont {D.}~\bibnamefont
  {Poulin}}\ and\ \bibinfo {author} {\bibfnamefont {P.}~\bibnamefont
  {Wocjan}},\ }\bibfield  {title} {\bibinfo {title} {Preparing ground states of
  quantum many-body systems on a quantum computer},\ }\href@noop {} {\bibfield
  {journal} {\bibinfo  {journal} {Physical review letters}\ }\textbf {\bibinfo
  {volume} {102}},\ \bibinfo {pages} {130503} (\bibinfo {year}
  {2009})}\BibitemShut {NoStop}%
\bibitem [{\citenamefont {Peruzzo}\ \emph {et~al.}(2014)\citenamefont
  {Peruzzo}, \citenamefont {McClean}, \citenamefont {Shadbolt}, \citenamefont
  {Yung}, \citenamefont {Zhou}, \citenamefont {Love}, \citenamefont
  {Aspuru-Guzik},\ and\ \citenamefont {Obrien}}]{Peruzzo.14}%
  \BibitemOpen
  \bibfield  {author} {\bibinfo {author} {\bibfnamefont {A.}~\bibnamefont
  {Peruzzo}}, \bibinfo {author} {\bibfnamefont {J.}~\bibnamefont {McClean}},
  \bibinfo {author} {\bibfnamefont {P.}~\bibnamefont {Shadbolt}}, \bibinfo
  {author} {\bibfnamefont {M.-H.}\ \bibnamefont {Yung}}, \bibinfo {author}
  {\bibfnamefont {X.-Q.}\ \bibnamefont {Zhou}}, \bibinfo {author}
  {\bibfnamefont {P.~J.}\ \bibnamefont {Love}}, \bibinfo {author}
  {\bibfnamefont {A.}~\bibnamefont {Aspuru-Guzik}}, \ and\ \bibinfo {author}
  {\bibfnamefont {J.~L.}\ \bibnamefont {Obrien}},\ }\bibfield  {title}
  {\bibinfo {title} {A variational eigenvalue solver on a photonic quantum
  processor},\ }\href@noop {} {\bibfield  {journal} {\bibinfo  {journal}
  {Nature communications}\ }\textbf {\bibinfo {volume} {5}},\ \bibinfo {pages}
  {4213} (\bibinfo {year} {2014})}\BibitemShut {NoStop}%
\bibitem [{\citenamefont {Somma}(2019)}]{Somma.19}%
  \BibitemOpen
  \bibfield  {author} {\bibinfo {author} {\bibfnamefont {R.~D.}\ \bibnamefont
  {Somma}},\ }\bibfield  {title} {\bibinfo {title} {Quantum eigenvalue
  estimation via time series analysis},\ }\href@noop {} {\bibfield  {journal}
  {\bibinfo  {journal} {New Journal of Physics}\ }\textbf {\bibinfo {volume}
  {21}},\ \bibinfo {pages} {123025} (\bibinfo {year} {2019})}\BibitemShut
  {NoStop}%
\bibitem [{\citenamefont {Ge}\ \emph {et~al.}(2019)\citenamefont {Ge},
  \citenamefont {Tura},\ and\ \citenamefont {Cirac}}]{Ge.19}%
  \BibitemOpen
  \bibfield  {author} {\bibinfo {author} {\bibfnamefont {Y.}~\bibnamefont
  {Ge}}, \bibinfo {author} {\bibfnamefont {J.}~\bibnamefont {Tura}}, \ and\
  \bibinfo {author} {\bibfnamefont {J.~I.}\ \bibnamefont {Cirac}},\ }\bibfield
  {title} {\bibinfo {title} {Faster ground state preparation and high-precision
  ground energy estimation with fewer qubits},\ }\href@noop {} {\bibfield
  {journal} {\bibinfo  {journal} {Journal of Mathematical Physics}\ }\textbf
  {\bibinfo {volume} {60}} (\bibinfo {year} {2019})}\BibitemShut {NoStop}%
\bibitem [{\citenamefont {Lin}\ and\ \citenamefont {Tong}(2020)}]{Lin.20}%
  \BibitemOpen
  \bibfield  {author} {\bibinfo {author} {\bibfnamefont {L.}~\bibnamefont
  {Lin}}\ and\ \bibinfo {author} {\bibfnamefont {Y.}~\bibnamefont {Tong}},\
  }\bibfield  {title} {\bibinfo {title} {Near-optimal ground state
  preparation},\ }\href@noop {} {\bibfield  {journal} {\bibinfo  {journal}
  {Quantum}\ }\textbf {\bibinfo {volume} {4}},\ \bibinfo {pages} {372}
  (\bibinfo {year} {2020})}\BibitemShut {NoStop}%
\bibitem [{\citenamefont {Zeng}\ \emph {et~al.}(2021)\citenamefont {Zeng},
  \citenamefont {Sun},\ and\ \citenamefont {Yuan}}]{Zeng.21}%
  \BibitemOpen
  \bibfield  {author} {\bibinfo {author} {\bibfnamefont {P.}~\bibnamefont
  {Zeng}}, \bibinfo {author} {\bibfnamefont {J.}~\bibnamefont {Sun}}, \ and\
  \bibinfo {author} {\bibfnamefont {X.}~\bibnamefont {Yuan}},\ }\bibfield
  {title} {\bibinfo {title} {Universal quantum algorithmic cooling on a quantum
  computer},\ }\href@noop {} {\bibfield  {journal} {\bibinfo  {journal} {arXiv
  preprint arXiv:2109.15304}\ } (\bibinfo {year} {2021})}\BibitemShut {NoStop}%
\bibitem [{\citenamefont {Gharibian}\ and\ \citenamefont
  {Le~Gall}(2022)}]{Gharibian.22}%
  \BibitemOpen
  \bibfield  {author} {\bibinfo {author} {\bibfnamefont {S.}~\bibnamefont
  {Gharibian}}\ and\ \bibinfo {author} {\bibfnamefont {F.}~\bibnamefont
  {Le~Gall}},\ }\bibfield  {title} {\bibinfo {title} {Dequantizing the quantum
  singular value transformation: hardness and applications to quantum chemistry
  and the quantum pcp conjecture},\ }in\ \href@noop {} {\emph {\bibinfo
  {booktitle} {Proceedings of the 54th Annual ACM SIGACT Symposium on Theory of
  Computing}}}\ (\bibinfo {year} {2022})\ pp.\ \bibinfo {pages}
  {19--32}\BibitemShut {NoStop}%
\bibitem [{\citenamefont {Albash}\ and\ \citenamefont
  {Lidar}(2018)}]{Albash.18}%
  \BibitemOpen
  \bibfield  {author} {\bibinfo {author} {\bibfnamefont {T.}~\bibnamefont
  {Albash}}\ and\ \bibinfo {author} {\bibfnamefont {D.~A.}\ \bibnamefont
  {Lidar}},\ }\bibfield  {title} {\bibinfo {title} {Adiabatic quantum
  computation},\ }\href@noop {} {\bibfield  {journal} {\bibinfo  {journal}
  {Reviews of Modern Physics}\ }\textbf {\bibinfo {volume} {90}},\ \bibinfo
  {pages} {015002} (\bibinfo {year} {2018})}\BibitemShut {NoStop}%
\bibitem [{\citenamefont {Shao}(2019)}]{Shao.19}%
  \BibitemOpen
  \bibfield  {author} {\bibinfo {author} {\bibfnamefont {C.}~\bibnamefont
  {Shao}},\ }\bibfield  {title} {\bibinfo {title} {Computing eigenvalues of
  matrices in a quantum computer},\ }\href@noop {} {\bibfield  {journal}
  {\bibinfo  {journal} {arXiv preprint arXiv:1912.08015}\ } (\bibinfo {year}
  {2019})}\BibitemShut {NoStop}%
\bibitem [{\citenamefont {Endo}\ \emph {et~al.}(2020)\citenamefont {Endo},
  \citenamefont {Sun}, \citenamefont {Li}, \citenamefont {Benjamin},\ and\
  \citenamefont {Yuan}}]{Endo.20}%
  \BibitemOpen
  \bibfield  {author} {\bibinfo {author} {\bibfnamefont {S.}~\bibnamefont
  {Endo}}, \bibinfo {author} {\bibfnamefont {J.}~\bibnamefont {Sun}}, \bibinfo
  {author} {\bibfnamefont {Y.}~\bibnamefont {Li}}, \bibinfo {author}
  {\bibfnamefont {S.~C.}\ \bibnamefont {Benjamin}}, \ and\ \bibinfo {author}
  {\bibfnamefont {X.}~\bibnamefont {Yuan}},\ }\bibfield  {title} {\bibinfo
  {title} {Variational quantum simulation of general processes},\ }\href@noop
  {} {\bibfield  {journal} {\bibinfo  {journal} {Physical Review Letters}\
  }\textbf {\bibinfo {volume} {125}},\ \bibinfo {pages} {010501} (\bibinfo
  {year} {2020})}\BibitemShut {NoStop}%
\bibitem [{\citenamefont {Yoshioka}\ \emph {et~al.}(2020)\citenamefont
  {Yoshioka}, \citenamefont {Nakagawa}, \citenamefont {Mitarai},\ and\
  \citenamefont {Fujii}}]{Yoshioka.20}%
  \BibitemOpen
  \bibfield  {author} {\bibinfo {author} {\bibfnamefont {N.}~\bibnamefont
  {Yoshioka}}, \bibinfo {author} {\bibfnamefont {Y.~O.}\ \bibnamefont
  {Nakagawa}}, \bibinfo {author} {\bibfnamefont {K.}~\bibnamefont {Mitarai}}, \
  and\ \bibinfo {author} {\bibfnamefont {K.}~\bibnamefont {Fujii}},\ }\bibfield
   {title} {\bibinfo {title} {Variational quantum algorithm for nonequilibrium
  steady states},\ }\href@noop {} {\bibfield  {journal} {\bibinfo  {journal}
  {Physical Review Research}\ }\textbf {\bibinfo {volume} {2}},\ \bibinfo
  {pages} {043289} (\bibinfo {year} {2020})}\BibitemShut {NoStop}%
\bibitem [{\citenamefont {Xie}\ \emph {et~al.}(2023)\citenamefont {Xie},
  \citenamefont {Xue},\ and\ \citenamefont {Zhang}}]{Xie.23}%
  \BibitemOpen
  \bibfield  {author} {\bibinfo {author} {\bibfnamefont {X.-D.}\ \bibnamefont
  {Xie}}, \bibinfo {author} {\bibfnamefont {Z.-Y.}\ \bibnamefont {Xue}}, \ and\
  \bibinfo {author} {\bibfnamefont {D.-B.}\ \bibnamefont {Zhang}},\ }\bibfield
  {title} {\bibinfo {title} {Variational quantum eigensolvers for the
  non-hermitian systems by variance minimization},\ }\href@noop {} {\bibfield
  {journal} {\bibinfo  {journal} {arXiv:2305.19807}\ } (\bibinfo {year}
  {2023})}\BibitemShut {NoStop}%
\bibitem [{\citenamefont {Zhao}\ \emph {et~al.}(2023)\citenamefont {Zhao},
  \citenamefont {Zhang},\ and\ \citenamefont {Wei}}]{zhao.23}%
  \BibitemOpen
  \bibfield  {author} {\bibinfo {author} {\bibfnamefont {H.}~\bibnamefont
  {Zhao}}, \bibinfo {author} {\bibfnamefont {P.}~\bibnamefont {Zhang}}, \ and\
  \bibinfo {author} {\bibfnamefont {T.-C.}\ \bibnamefont {Wei}},\ }\bibfield
  {title} {\bibinfo {title} {A universal variational quantum eigensolver for
  non-hermitian systems},\ }\href@noop {} {\bibfield  {journal} {\bibinfo
  {journal} {Scientific Reports}\ }\textbf {\bibinfo {volume} {13}},\ \bibinfo
  {pages} {22313} (\bibinfo {year} {2023})}\BibitemShut {NoStop}%
\bibitem [{\citenamefont {Low}\ and\ \citenamefont {Su}(2024)}]{Low.24}%
  \BibitemOpen
  \bibfield  {author} {\bibinfo {author} {\bibfnamefont {G.~H.}\ \bibnamefont
  {Low}}\ and\ \bibinfo {author} {\bibfnamefont {Y.}~\bibnamefont {Su}},\
  }\bibfield  {title} {\bibinfo {title} {Quantum eigenvalue processing},\
  }\href@noop {} {\bibfield  {journal} {\bibinfo  {journal} {arXiv:2401.06240}\
  } (\bibinfo {year} {2024})}\BibitemShut {NoStop}%
\bibitem [{\citenamefont {Alase}\ and\ \citenamefont
  {Karuvade}(2024)}]{Alase.24}%
  \BibitemOpen
  \bibfield  {author} {\bibinfo {author} {\bibfnamefont {A.}~\bibnamefont
  {Alase}}\ and\ \bibinfo {author} {\bibfnamefont {S.}~\bibnamefont
  {Karuvade}},\ }\bibfield  {title} {\bibinfo {title} {Resolvent-based quantum
  phase estimation: Towards estimation of parametrized eigenvalues},\
  }\href@noop {} {\bibfield  {journal} {\bibinfo  {journal} {arXiv preprint
  arXiv:2410.04837}\ } (\bibinfo {year} {2024})}\BibitemShut {NoStop}%
\bibitem [{\citenamefont {Kawabata}\ \emph {et~al.}(2019)\citenamefont
  {Kawabata}, \citenamefont {Shiozaki}, \citenamefont {Ueda},\ and\
  \citenamefont {Sato}}]{Kawabata.19}%
  \BibitemOpen
  \bibfield  {author} {\bibinfo {author} {\bibfnamefont {K.}~\bibnamefont
  {Kawabata}}, \bibinfo {author} {\bibfnamefont {K.}~\bibnamefont {Shiozaki}},
  \bibinfo {author} {\bibfnamefont {M.}~\bibnamefont {Ueda}}, \ and\ \bibinfo
  {author} {\bibfnamefont {M.}~\bibnamefont {Sato}},\ }\bibfield  {title}
  {\bibinfo {title} {Symmetry and topology in non-hermitian physics},\
  }\href@noop {} {\bibfield  {journal} {\bibinfo  {journal} {Physical Review
  X}\ }\textbf {\bibinfo {volume} {9}},\ \bibinfo {pages} {041015} (\bibinfo
  {year} {2019})}\BibitemShut {NoStop}%
\bibitem [{\citenamefont {Borgnia}\ \emph {et~al.}(2020)\citenamefont
  {Borgnia}, \citenamefont {Kruchkov},\ and\ \citenamefont
  {Slager}}]{Borgnia.20}%
  \BibitemOpen
  \bibfield  {author} {\bibinfo {author} {\bibfnamefont {D.~S.}\ \bibnamefont
  {Borgnia}}, \bibinfo {author} {\bibfnamefont {A.~J.}\ \bibnamefont
  {Kruchkov}}, \ and\ \bibinfo {author} {\bibfnamefont {R.-J.}\ \bibnamefont
  {Slager}},\ }\bibfield  {title} {\bibinfo {title} {Non-hermitian boundary
  modes and topology},\ }\href@noop {} {\bibfield  {journal} {\bibinfo
  {journal} {Physical review letters}\ }\textbf {\bibinfo {volume} {124}},\
  \bibinfo {pages} {056802} (\bibinfo {year} {2020})}\BibitemShut {NoStop}%
\bibitem [{\citenamefont {Bergholtz}\ \emph {et~al.}(2021)\citenamefont
  {Bergholtz}, \citenamefont {Budich},\ and\ \citenamefont
  {Kunst}}]{Bergholtz.21}%
  \BibitemOpen
  \bibfield  {author} {\bibinfo {author} {\bibfnamefont {E.~J.}\ \bibnamefont
  {Bergholtz}}, \bibinfo {author} {\bibfnamefont {J.~C.}\ \bibnamefont
  {Budich}}, \ and\ \bibinfo {author} {\bibfnamefont {F.~K.}\ \bibnamefont
  {Kunst}},\ }\bibfield  {title} {\bibinfo {title} {Exceptional topology of
  non-hermitian systems},\ }\href@noop {} {\bibfield  {journal} {\bibinfo
  {journal} {Reviews of Modern Physics}\ }\textbf {\bibinfo {volume} {93}},\
  \bibinfo {pages} {015005} (\bibinfo {year} {2021})}\BibitemShut {NoStop}%
\bibitem [{\citenamefont {Medvedyeva}\ \emph {et~al.}(2016)\citenamefont
  {Medvedyeva}, \citenamefont {Essler},\ and\ \citenamefont
  {Prosen}}]{Medvedyeva.16}%
  \BibitemOpen
  \bibfield  {author} {\bibinfo {author} {\bibfnamefont {M.~V.}\ \bibnamefont
  {Medvedyeva}}, \bibinfo {author} {\bibfnamefont {F.~H.}\ \bibnamefont
  {Essler}}, \ and\ \bibinfo {author} {\bibfnamefont {T.}~\bibnamefont
  {Prosen}},\ }\bibfield  {title} {\bibinfo {title} {Exact bethe ansatz
  spectrum of a tight-binding chain with dephasing noise},\ }\href@noop {}
  {\bibfield  {journal} {\bibinfo  {journal} {Physical review letters}\
  }\textbf {\bibinfo {volume} {117}},\ \bibinfo {pages} {137202} (\bibinfo
  {year} {2016})}\BibitemShut {NoStop}%
\bibitem [{\citenamefont {Banchi}\ \emph {et~al.}(2017)\citenamefont {Banchi},
  \citenamefont {Burgarth},\ and\ \citenamefont {Kastoryano}}]{Banchi.17}%
  \BibitemOpen
  \bibfield  {author} {\bibinfo {author} {\bibfnamefont {L.}~\bibnamefont
  {Banchi}}, \bibinfo {author} {\bibfnamefont {D.}~\bibnamefont {Burgarth}}, \
  and\ \bibinfo {author} {\bibfnamefont {M.~J.}\ \bibnamefont {Kastoryano}},\
  }\bibfield  {title} {\bibinfo {title} {Driven quantum dynamics: Will it
  blend?}\ }\href@noop {} {\bibfield  {journal} {\bibinfo  {journal} {Physical
  Review X}\ }\textbf {\bibinfo {volume} {7}},\ \bibinfo {pages} {041015}
  (\bibinfo {year} {2017})}\BibitemShut {NoStop}%
\bibitem [{\citenamefont {Rowlands}\ and\ \citenamefont
  {Lamacraft}(2018)}]{Rowlands.18}%
  \BibitemOpen
  \bibfield  {author} {\bibinfo {author} {\bibfnamefont {D.~A.}\ \bibnamefont
  {Rowlands}}\ and\ \bibinfo {author} {\bibfnamefont {A.}~\bibnamefont
  {Lamacraft}},\ }\bibfield  {title} {\bibinfo {title} {Noisy spins and the
  richardson-gaudin model},\ }\href@noop {} {\bibfield  {journal} {\bibinfo
  {journal} {Physical review letters}\ }\textbf {\bibinfo {volume} {120}},\
  \bibinfo {pages} {090401} (\bibinfo {year} {2018})}\BibitemShut {NoStop}%
\bibitem [{\citenamefont {Mori}\ and\ \citenamefont {Shirai}(2020)}]{Mori.20}%
  \BibitemOpen
  \bibfield  {author} {\bibinfo {author} {\bibfnamefont {T.}~\bibnamefont
  {Mori}}\ and\ \bibinfo {author} {\bibfnamefont {T.}~\bibnamefont {Shirai}},\
  }\bibfield  {title} {\bibinfo {title} {Resolving a discrepancy between
  liouvillian gap and relaxation time in boundary-dissipated quantum many-body
  systems},\ }\href@noop {} {\bibfield  {journal} {\bibinfo  {journal}
  {Physical Review Letters}\ }\textbf {\bibinfo {volume} {125}},\ \bibinfo
  {pages} {230604} (\bibinfo {year} {2020})}\BibitemShut {NoStop}%
\bibitem [{\citenamefont {Yuan}\ \emph {et~al.}(2021)\citenamefont {Yuan},
  \citenamefont {Wang}, \citenamefont {Wang},\ and\ \citenamefont
  {Deng}}]{Yuan.21}%
  \BibitemOpen
  \bibfield  {author} {\bibinfo {author} {\bibfnamefont {D.}~\bibnamefont
  {Yuan}}, \bibinfo {author} {\bibfnamefont {H.-R.}\ \bibnamefont {Wang}},
  \bibinfo {author} {\bibfnamefont {Z.}~\bibnamefont {Wang}}, \ and\ \bibinfo
  {author} {\bibfnamefont {D.-L.}\ \bibnamefont {Deng}},\ }\bibfield  {title}
  {\bibinfo {title} {Solving the liouvillian gap with artificial neural
  networks},\ }\href@noop {} {\bibfield  {journal} {\bibinfo  {journal}
  {Physical Review Letters}\ }\textbf {\bibinfo {volume} {126}},\ \bibinfo
  {pages} {160401} (\bibinfo {year} {2021})}\BibitemShut {NoStop}%
\bibitem [{\citenamefont {Zhou}\ \emph {et~al.}(2022)\citenamefont {Zhou},
  \citenamefont {Wang},\ and\ \citenamefont {Chen}}]{Zhou.22}%
  \BibitemOpen
  \bibfield  {author} {\bibinfo {author} {\bibfnamefont {B.}~\bibnamefont
  {Zhou}}, \bibinfo {author} {\bibfnamefont {X.}~\bibnamefont {Wang}}, \ and\
  \bibinfo {author} {\bibfnamefont {S.}~\bibnamefont {Chen}},\ }\bibfield
  {title} {\bibinfo {title} {Exponential size scaling of the liouvillian gap in
  boundary-dissipated systems with anderson localization},\ }\href@noop {}
  {\bibfield  {journal} {\bibinfo  {journal} {Physical Review B}\ }\textbf
  {\bibinfo {volume} {106}},\ \bibinfo {pages} {064203} (\bibinfo {year}
  {2022})}\BibitemShut {NoStop}%
\bibitem [{sm()}]{sm}%
  \BibitemOpen
  \href@noop {} {}\bibinfo {note} {See Supplemental Material.}\BibitemShut
  {Stop}%
\bibitem [{\citenamefont {Horn}\ and\ \citenamefont {Johnson}(2012)}]{Horn.12}%
  \BibitemOpen
  \bibfield  {author} {\bibinfo {author} {\bibfnamefont {R.~A.}\ \bibnamefont
  {Horn}}\ and\ \bibinfo {author} {\bibfnamefont {C.~R.}\ \bibnamefont
  {Johnson}},\ }\href@noop {} {\emph {\bibinfo {title} {Matrix analysis}}}\
  (\bibinfo  {publisher} {Cambridge university press},\ \bibinfo {year}
  {2012})\BibitemShut {NoStop}%
\bibitem [{\citenamefont {Gily{\'e}n}\ \emph {et~al.}(2019)\citenamefont
  {Gily{\'e}n}, \citenamefont {Su}, \citenamefont {Low},\ and\ \citenamefont
  {Wiebe}}]{Gilyen.19}%
  \BibitemOpen
  \bibfield  {author} {\bibinfo {author} {\bibfnamefont {A.}~\bibnamefont
  {Gily{\'e}n}}, \bibinfo {author} {\bibfnamefont {Y.}~\bibnamefont {Su}},
  \bibinfo {author} {\bibfnamefont {G.~H.}\ \bibnamefont {Low}}, \ and\
  \bibinfo {author} {\bibfnamefont {N.}~\bibnamefont {Wiebe}},\ }\bibfield
  {title} {\bibinfo {title} {Quantum singular value transformation and beyond:
  exponential improvements for quantum matrix arithmetics},\ }in\ \href@noop {}
  {\emph {\bibinfo {booktitle} {Proceedings of the 51st Annual ACM SIGACT
  Symposium on Theory of Computing}}}\ (\bibinfo {year} {2019})\ pp.\ \bibinfo
  {pages} {193--204}\BibitemShut {NoStop}%
\bibitem [{\citenamefont {Childs}\ and\ \citenamefont
  {Wiebe}(2012)}]{Childs.12}%
  \BibitemOpen
  \bibfield  {author} {\bibinfo {author} {\bibfnamefont {A.~M.}\ \bibnamefont
  {Childs}}\ and\ \bibinfo {author} {\bibfnamefont {N.}~\bibnamefont {Wiebe}},\
  }\bibfield  {title} {\bibinfo {title} {Hamiltonian simulation using linear
  combinations of unitary operations},\ }\href@noop {} {\bibfield  {journal}
  {\bibinfo  {journal} {arXiv:1202.5822}\ } (\bibinfo {year}
  {2012})}\BibitemShut {NoStop}%
\bibitem [{\citenamefont {Nielsen}\ and\ \citenamefont
  {Chuang}(2000)}]{Nielsen.02}%
  \BibitemOpen
  \bibfield  {author} {\bibinfo {author} {\bibfnamefont {M.~A.}\ \bibnamefont
  {Nielsen}}\ and\ \bibinfo {author} {\bibfnamefont {I.}~\bibnamefont
  {Chuang}},\ }\href@noop {} {\emph {\bibinfo {title} {Quantum computation and
  quantum information}}}\ (\bibinfo  {publisher} {Cambridge University Press,
  Cambridge},\ \bibinfo {year} {2000})\BibitemShut {NoStop}%
\bibitem [{\citenamefont {Dong}\ \emph {et~al.}(2022)\citenamefont {Dong},
  \citenamefont {Lin},\ and\ \citenamefont {Tong}}]{Dong.22}%
  \BibitemOpen
  \bibfield  {author} {\bibinfo {author} {\bibfnamefont {Y.}~\bibnamefont
  {Dong}}, \bibinfo {author} {\bibfnamefont {L.}~\bibnamefont {Lin}}, \ and\
  \bibinfo {author} {\bibfnamefont {Y.}~\bibnamefont {Tong}},\ }\bibfield
  {title} {\bibinfo {title} {Ground-state preparation and energy estimation on
  early fault-tolerant quantum computers via quantum eigenvalue transformation
  of unitary matrices},\ }\href@noop {} {\bibfield  {journal} {\bibinfo
  {journal} {PRX Quantum}\ }\textbf {\bibinfo {volume} {3}},\ \bibinfo {pages}
  {040305} (\bibinfo {year} {2022})}\BibitemShut {NoStop}%
\bibitem [{\citenamefont {Low}(2017)}]{low2017}%
  \BibitemOpen
  \bibfield  {author} {\bibinfo {author} {\bibfnamefont {G.~H.}\ \bibnamefont
  {Low}},\ }\emph {\bibinfo {title} {Quantum signal processing by single-qubit
  dynamics}},\ \href@noop {} {Ph.D. thesis},\ \bibinfo  {school} {Massachusetts
  Institute of Technology} (\bibinfo {year} {2017})\BibitemShut {NoStop}%
\bibitem [{\citenamefont {Ding}\ \emph {et~al.}(2024)\citenamefont {Ding},
  \citenamefont {Chen},\ and\ \citenamefont {Lin}}]{ding2024single}%
  \BibitemOpen
  \bibfield  {author} {\bibinfo {author} {\bibfnamefont {Z.}~\bibnamefont
  {Ding}}, \bibinfo {author} {\bibfnamefont {C.-F.}\ \bibnamefont {Chen}}, \
  and\ \bibinfo {author} {\bibfnamefont {L.}~\bibnamefont {Lin}},\ }\bibfield
  {title} {\bibinfo {title} {Single-ancilla ground state preparation via
  lindbladians},\ }\href@noop {} {\bibfield  {journal} {\bibinfo  {journal}
  {Physical Review Research}\ }\textbf {\bibinfo {volume} {6}},\ \bibinfo
  {pages} {033147} (\bibinfo {year} {2024})}\BibitemShut {NoStop}%
\bibitem [{\citenamefont {Lin}(2025)}]{lin2025dissipative}%
  \BibitemOpen
  \bibfield  {author} {\bibinfo {author} {\bibfnamefont {L.}~\bibnamefont
  {Lin}},\ }\bibfield  {title} {\bibinfo {title} {Dissipative preparation of
  many-body quantum states: Towards practical quantum advantage},\ }\href@noop
  {} {\bibfield  {journal} {\bibinfo  {journal} {arXiv preprint
  arXiv:2505.21308}\ } (\bibinfo {year} {2025})}\BibitemShut {NoStop}%
\bibitem [{\citenamefont {Zhan}\ \emph {et~al.}(2025)\citenamefont {Zhan},
  \citenamefont {Ding}, \citenamefont {Huhn}, \citenamefont {Gray},
  \citenamefont {Preskill}, \citenamefont {Chan},\ and\ \citenamefont
  {Lin}}]{zhan2025rapid}%
  \BibitemOpen
  \bibfield  {author} {\bibinfo {author} {\bibfnamefont {Y.}~\bibnamefont
  {Zhan}}, \bibinfo {author} {\bibfnamefont {Z.}~\bibnamefont {Ding}}, \bibinfo
  {author} {\bibfnamefont {J.}~\bibnamefont {Huhn}}, \bibinfo {author}
  {\bibfnamefont {J.}~\bibnamefont {Gray}}, \bibinfo {author} {\bibfnamefont
  {J.}~\bibnamefont {Preskill}}, \bibinfo {author} {\bibfnamefont {G.~K.}\
  \bibnamefont {Chan}}, \ and\ \bibinfo {author} {\bibfnamefont
  {L.}~\bibnamefont {Lin}},\ }\bibfield  {title} {\bibinfo {title} {Rapid
  quantum ground state preparation via dissipative dynamics},\ }\href@noop {}
  {\bibfield  {journal} {\bibinfo  {journal} {arXiv preprint arXiv:2503.15827}\
  } (\bibinfo {year} {2025})}\BibitemShut {NoStop}%
\bibitem [{\citenamefont {Norris}(1998)}]{Norris.98}%
  \BibitemOpen
  \bibfield  {author} {\bibinfo {author} {\bibfnamefont {J.~R.}\ \bibnamefont
  {Norris}},\ }\href@noop {} {\emph {\bibinfo {title} {Markov chains}}},\
  \bibinfo {number} {2}\ (\bibinfo  {publisher} {Cambridge university press},\
  \bibinfo {year} {1998})\BibitemShut {NoStop}%
\bibitem [{\citenamefont {Odencrantz}(2000)}]{Odencrantz.00}%
  \BibitemOpen
  \bibfield  {author} {\bibinfo {author} {\bibfnamefont {J.}~\bibnamefont
  {Odencrantz}},\ }\bibfield  {title} {\bibinfo {title} {Markov chains: Gibbs
  fields, monte carlo simulation, and queues},\ }\href@noop {} {\bibfield
  {journal} {\bibinfo  {journal} {Technometrics}\ }\textbf {\bibinfo {volume}
  {42}},\ \bibinfo {pages} {438} (\bibinfo {year} {2000})}\BibitemShut
  {NoStop}%
\bibitem [{\citenamefont {Ibe}(2013)}]{Ibe.12}%
  \BibitemOpen
  \bibfield  {author} {\bibinfo {author} {\bibfnamefont {O.}~\bibnamefont
  {Ibe}},\ }\href@noop {} {\emph {\bibinfo {title} {Markov processes for
  stochastic modeling}}}\ (\bibinfo  {publisher} {Newnes},\ \bibinfo {year}
  {2013})\BibitemShut {NoStop}%
\bibitem [{\citenamefont {Meyn}\ and\ \citenamefont {Tweedie}(2012)}]{Meyn.12}%
  \BibitemOpen
  \bibfield  {author} {\bibinfo {author} {\bibfnamefont {S.~P.}\ \bibnamefont
  {Meyn}}\ and\ \bibinfo {author} {\bibfnamefont {R.~L.}\ \bibnamefont
  {Tweedie}},\ }\href@noop {} {\emph {\bibinfo {title} {Markov chains and
  stochastic stability}}}\ (\bibinfo  {publisher} {Springer Science \& Business
  Media},\ \bibinfo {year} {2012})\BibitemShut {NoStop}%
\bibitem [{\citenamefont {Levin}\ and\ \citenamefont {Peres}(2017)}]{Levin.17}%
  \BibitemOpen
  \bibfield  {author} {\bibinfo {author} {\bibfnamefont {D.~A.}\ \bibnamefont
  {Levin}}\ and\ \bibinfo {author} {\bibfnamefont {Y.}~\bibnamefont {Peres}},\
  }\href@noop {} {\emph {\bibinfo {title} {Markov chains and mixing times}}},\
  Vol.\ \bibinfo {volume} {107}\ (\bibinfo  {publisher} {American Mathematical
  Soc.},\ \bibinfo {year} {2017})\BibitemShut {NoStop}%
\bibitem [{\citenamefont {Kahan}\ \emph {et~al.}(1982)\citenamefont {Kahan},
  \citenamefont {Parlett},\ and\ \citenamefont {Jiang}}]{Kahan.82}%
  \BibitemOpen
  \bibfield  {author} {\bibinfo {author} {\bibfnamefont {W.}~\bibnamefont
  {Kahan}}, \bibinfo {author} {\bibfnamefont {B.}~\bibnamefont {Parlett}}, \
  and\ \bibinfo {author} {\bibfnamefont {E.}~\bibnamefont {Jiang}},\ }\bibfield
   {title} {\bibinfo {title} {Residual bounds on approximate eigensystems of
  nonnormal matrices},\ }\href@noop {} {\bibfield  {journal} {\bibinfo
  {journal} {SIAM Journal on Numerical Analysis}\ }\textbf {\bibinfo {volume}
  {19}},\ \bibinfo {pages} {470} (\bibinfo {year} {1982})}\BibitemShut
  {NoStop}%
\bibitem [{\citenamefont {Erxiong}(1994)}]{Jiang.94}%
  \BibitemOpen
  \bibfield  {author} {\bibinfo {author} {\bibfnamefont {J.}~\bibnamefont
  {Erxiong}},\ }\bibfield  {title} {\bibinfo {title} {Bounds for the smallest
  singular value of a jordan block with an application to eigenvalue
  perturbation},\ }\href@noop {} {\bibfield  {journal} {\bibinfo  {journal}
  {Linear Algebra and its Applications}\ }\textbf {\bibinfo {volume} {197}},\
  \bibinfo {pages} {691} (\bibinfo {year} {1994})}\BibitemShut {NoStop}%
\bibitem [{\citenamefont {Breuer}\ and\ \citenamefont
  {Petruccione}(2002)}]{Breuer.02}%
  \BibitemOpen
  \bibfield  {author} {\bibinfo {author} {\bibfnamefont {H.-P.}\ \bibnamefont
  {Breuer}}\ and\ \bibinfo {author} {\bibfnamefont {F.}~\bibnamefont
  {Petruccione}},\ }\href@noop {} {\emph {\bibinfo {title} {The theory of open
  quantum systems}}}\ (\bibinfo  {publisher} {Oxford University Press, USA},\
  \bibinfo {year} {2002})\BibitemShut {NoStop}%
\bibitem [{\citenamefont {Low}\ and\ \citenamefont {Chuang}(2019)}]{Low.19}%
  \BibitemOpen
  \bibfield  {author} {\bibinfo {author} {\bibfnamefont {G.~H.}\ \bibnamefont
  {Low}}\ and\ \bibinfo {author} {\bibfnamefont {I.~L.}\ \bibnamefont
  {Chuang}},\ }\bibfield  {title} {\bibinfo {title} {Hamiltonian simulation by
  qubitization},\ }\href@noop {} {\bibfield  {journal} {\bibinfo  {journal}
  {Quantum}\ }\textbf {\bibinfo {volume} {3}},\ \bibinfo {pages} {163}
  (\bibinfo {year} {2019})}\BibitemShut {NoStop}%
\bibitem [{\citenamefont {Zhang}\ and\ \citenamefont {Yuan}(2024)}]{Zhang.24}%
  \BibitemOpen
  \bibfield  {author} {\bibinfo {author} {\bibfnamefont {X.-M.}\ \bibnamefont
  {Zhang}}\ and\ \bibinfo {author} {\bibfnamefont {X.}~\bibnamefont {Yuan}},\
  }\bibfield  {title} {\bibinfo {title} {Circuit complexity of quantum access
  models for encoding classical data},\ }\href@noop {} {\bibfield  {journal}
  {\bibinfo  {journal} {npj Quantum Information}\ }\textbf {\bibinfo {volume}
  {10}},\ \bibinfo {pages} {42} (\bibinfo {year} {2024})}\BibitemShut {NoStop}%
\bibitem [{\citenamefont {Li}\ \emph {et~al.}(2022)\citenamefont {Li},
  \citenamefont {Zheng}, \citenamefont {Gao},\ and\ \citenamefont
  {Long}}]{Li.22}%
  \BibitemOpen
  \bibfield  {author} {\bibinfo {author} {\bibfnamefont {X.}~\bibnamefont
  {Li}}, \bibinfo {author} {\bibfnamefont {C.}~\bibnamefont {Zheng}}, \bibinfo
  {author} {\bibfnamefont {J.}~\bibnamefont {Gao}}, \ and\ \bibinfo {author}
  {\bibfnamefont {G.}~\bibnamefont {Long}},\ }\href@noop {} {\bibinfo {title}
  {Dynamics simulation and numerical analysis of arbitrary time-dependent
  $\mathcal{PT}$-symmetric system based on density operators},\ } (\bibinfo
  {year} {2022})\BibitemShut {NoStop}%
\bibitem [{\citenamefont {Melkani}(2023)}]{Melkani.23}%
  \BibitemOpen
  \bibfield  {author} {\bibinfo {author} {\bibfnamefont {A.}~\bibnamefont
  {Melkani}},\ }\bibfield  {title} {\bibinfo {title} {Degeneracies and symmetry
  breaking in pseudo-hermitian matrices},\ }\href@noop {} {\bibfield  {journal}
  {\bibinfo  {journal} {Physical Review Research}\ }\textbf {\bibinfo {volume}
  {5}},\ \bibinfo {pages} {023035} (\bibinfo {year} {2023})}\BibitemShut
  {NoStop}%
\bibitem [{\citenamefont {Horn}\ and\ \citenamefont {Johnson}(1994)}]{Horn.94}%
  \BibitemOpen
  \bibfield  {author} {\bibinfo {author} {\bibfnamefont {R.~A.}\ \bibnamefont
  {Horn}}\ and\ \bibinfo {author} {\bibfnamefont {C.~R.}\ \bibnamefont
  {Johnson}},\ }\href@noop {} {\emph {\bibinfo {title} {Topics in matrix
  analysis}}}\ (\bibinfo  {publisher} {Cambridge university press},\ \bibinfo
  {year} {1994})\BibitemShut {NoStop}%
\end{thebibliography}

\end{document}